\documentclass{elsart1p}

\usepackage{graphicx}

\usepackage{amssymb,longtable,amsmath}

\begin{document}

\begin{frontmatter}



\title{Charged-Particle Thermonuclear Reaction Rates:\\ III. Nuclear Physics Input}


\author{C. Iliadis}, \author{R. Longland}, \author{A. E. Champagne}
\address{Department of Physics and Astronomy, University of North Carolina, Chapel Hill, NC 27599-3255, USA; Triangle Universities Nuclear Laboratory, Durham, NC 27708-0308, USA}
\author{A. Coc}
\address{Centre de Spectrom\'etrie Nucl\'eaire et de Spectrom\'etrie de Masse (CSNSM), UMR 8609, CNRS/IN2P3 and Universit\'e
        Paris Sud 11, B\^atiment 104, 91405 Orsay Campus, France}

\begin{abstract}
The nuclear physics input used to compute the Monte Carlo reaction rates and probability density functions that are tabulated in the second paper of this series (Paper II) is presented. Specifically, we publish the input files to the Monte Carlo reaction rate code \texttt{RatesMC}, which is based on the formalism presented in the first paper of this series  (Paper I). This data base contains overwhelmingly experimental nuclear physics information. The survey of literature for this review was concluded in November 2009. 
\end{abstract}


\end{frontmatter}

\section{Introduction}\label{intro}
For a number of reasons, the present evaluation of charged-particle thermonuclear reaction rates represents a significant step forward compared to previous work. First, we developed a new method of computing reaction rates, which is based on Monte Carlo techniques and assigns to each nuclear input quantity a physically motivated probability density function. The method is described in the first paper of this series (Paper I) and the numerical results for reaction rates and rate probability density functions are presented in the second paper of this series (Paper II). Second, a number of years have passed since the last two evaluations \cite{Ang99,Ili01} of similar scope as the present work have been published. The rapid progress seen in the field of nuclear astrophysics over the past few years clearly warrants a new charged-particle reaction rate evaluation. Third, thermonuclear reaction rates are not directly measured quantities, but are derived from a multitude of different nuclear physics input quantities. Consequently, the quality and reliability of a reaction rate evaluation hinges directly on the transparency and reproducibility of the input data. The last two aspects will be addressed in the present paper.

We present here the input files to the Monte Carlo code \texttt{RatesMC} containing the nuclear physics data used to compute our reaction rates. For each reaction we list nuclear properties; nonresonant S-factors; recommended resonance energies, strengths and partial widths; upper limits of resonance contributions; numerical rate integrations; and interferences between resonances of same spin and parity. Section 2 contains a brief discussion of general procedures, literature sources and the status of data. In Sec. 3 we focus attention on a number of important issues. A brief summary is given in Sec. 4. The meaning of each input row in a sample input file to \texttt{RatesMC} is explained in App. \ref{sec:inptfiles}. The actual nuclear physics input data for each of the reactions evaluated in the present work are listed in App. \ref{sec:nuclinput}. 

\section{Procedures, sources and status of data}

We will briefly describe our procedures for nuclear data analysis and evaluation. The details are too numerous to list here and depend on a case-by-case basis, but some general principles can be outlined. All the available sources of literature have been consulted. We focus our attention on refereed journals, but in exceptional circumstances conference proceedings and Ph.D. theses are also taken into account. If a particular quantity has been measured more than once, we adopt a weighted average (except for resonance strengths; see below), unless there was reason to exclude unreliable measurements. In some cases we succeeded in correcting original data for systematic effects, for example, improved stoichiometries, stopping powers, coincidence summing corrections, and so on. The reader may find the discussions in Refs. \cite{Ili08,Row02} illuminating.

For many resonances we use the evaluated energies of Endt \cite{End90,End98}. However, for papers published after 1998 we are compelled to perform our own evaluation. Resonance energies can be measured directly using thick-target excitation functions or are derived by using measured excitation energies and the reaction Q-value ($E_r^{cm}=E_x-Q$; Sec. 5.1.1 in Paper I). We adopted in general the method that resulted in the smallest uncertainties. For reaction Q-values we use the evaluated results of Audi and collaborators \cite{Aud03}, unless more recently (after 2003) measured masses have been reported in the literature (see Tab. 1 of Paper II). 

Assignments of nuclear level spins and parities are especially precarious in the recent literature. As pointed out in the introduction to his 1990 evaluation \cite{End90}, Endt carefully distinguished between strong (that is, model-independent) and weak arguments when assigning $J^{\pi}$ values. Assignments based on weak arguments were placed in parenthesis by Refs. \cite{End90,End98}. In most papers published after 1998 this important distinction between strong and weak arguments is blurred and it requires now significant efforts by a reviewer to disentangle the arguments for $J^{\pi}$ assignments from different reaction and decay studies. We can hardly overstate to our colleagues the importance of strictly following the established rules for assigning spins and parities (see introduction of Ref. \cite{End90} and references therein). 

Regarding resonance strengths, we did not follow in general the procedure of the NACRE reaction rate evaluation \cite{Ang99}, where for a given resonance the strength is found from a weighted mean of values obtained in different measurements. Instead, whenever possible we normalize literature results to a set of carefully measured ``standard" resonance strengths (see Tab. 4.12 in Iliadis \cite{Ili07}). Note that no standard resonance strengths exist at present, for example, for ($\alpha$,$\gamma$) reactions or for (p,$\gamma$) reactions on any of the Ne isotopes.

Partial widths can be derived from measured resonance strengths, mean lifetimes or spectroscopic factors. See Ref. \cite{Ili07} for details. We prefer, whenever feasible, to calculate reaction rates by numerical integration (see Eq. (1) of Paper I) using partial widths as input instead of computing them analytically using the resonance strength (see Eq. (10) of Paper I). The former procedure automatically accounts for the low- and high-energy tails of a resonance and makes any artificial corrections (for example, for ``nonresonant" S-factor tails) obsolete. In most cases involving short-lived targets, we compute the partial widths by using measured spectroscopic factors and mean lifetimes of the corresponding levels in the mirror nucleus. This method and its justification has been described in detail by Iliadis et al. \cite{Ili99}. 

Uncertainties are treated in the following manner. For measured resonance energies the reported or derived mean value and the corresponding (``1$\sigma$") uncertainty is associated with the parameters $\mu$ and $\sigma$, respectively, of a Gaussian probability density function (Sec. 5.1.1 of Paper I). For measured resonance strengths or partial widths we associate the mean value and corresponding uncertainty with the expectation value and the square root of the variance, respectively, of a lognormal distribution (Sec. 5.1.2 of Paper I). The lognormal parameters $\mu$ and $\sigma$ are then computed from Eq. (27) of Paper I. If uncertainties are not available from the literature, we use certain global values to the best of our judgement: (i) the direct capture reaction rate is usually calculated using the method described in Secs. 2.1 and 5.1.3 of Paper I. It is sometimes assumed that this procedure represents a purely theoretical approach. However, this assumption is incorrect since the absolute magnitude of the direct capture S-factor is determined by using {\it experimental} spectroscopic factors in Eq. (35) of Paper I. We adopt in most of these cases an uncertainty (square root of the variance) of 40\% for the direct capture S-factor (see Eq. (5) of Paper I). This value is based on a systematic comparison of experimental spectroscopic factors from direct capture and from transfer reaction studies \cite{Ili04}; (ii) when particle and $\gamma$-ray partial widths are calculated from measured spectroscopic factors and $\gamma$-ray transition strengths, we assume uncertainties of 40\% and 50\%, respectively. The choice of these values is supported by a systematic comparison of partial widths \cite{Ili99} and by an uncertainty analysis of measured spectroscopic factors \cite{Tho99}; (iii) in exceptional cases we adopt spectroscopic factors and $\gamma$-ray transition strengths from the nuclear shell model. For the sake of consistency, we assume values of 40\% and 50\% for the uncertainties of shell model based particle and $\gamma$-ray partial widths, respectively.

Upper limits of particle partial widths are sampled according to the procedure outlined in Sec. 5.2.1 of Paper I. Specifically, the Porter-Thomas distribution of dimensionless reduced widths is obtained with mean values of $\langle \theta^2_p \rangle=0.0045$ for protons (or neutrons) and $\langle \theta^2_{\alpha} \rangle=0.010$ for $\alpha$-particles, and the distribution is truncated at the experimental upper limit of the dimensionless reduced width (see Eq. (38) of Paper I). If the spin and parity of a particular nuclear level is unknown, we assume formation of the expected (but yet undetected) resonance via s-waves ($\ell=0$).

Table 1 of Paper II contains a list of references for each reaction evaluated in the present work. The list is not exhaustive by any means, but provides the reader with an impression on the most recent or relevant work considered here. We hope that making our recommended input data available to the community represents a step forward in terms of the reproducibility and transparency of evaluated reaction rates. 
 
\section{Words of caution}

The input data presented in App. \ref{sec:nuclinput} are based on the most reliable information that we are able to extract from the published literature. They reflect our best current knowledge of these parameters. By no means can we exclude the possibility that, for example, a reported resonance strength was derived using the wrong stoichiometry, or that an incorrect $J^{\pi}$ value has been reported for a particular nuclear level. This issue should be kept in mind when drawing conclusions from our Monte Carlo reaction rate uncertainties.

Many test runs have been performed to ensure proper functioning of our new Monte Carlo code \texttt{RatesMC}. It is gratifying to see that the present reaction rates agree with those calculated using the previous code \texttt{RateErrors} \cite{Tho99} in those simple and restricted circumstances where the latter code provides accurate results (see discussion in Sec. 3.3 of Paper I). However, a number of issues are disregarded by \texttt{RatesMC} although their implementation is straightforward from a computational point of view: (i)  when integrating the rate contribution of a resonance numerically, and the resonance can be formed via two orbital angular momenta, $\ell$ and $\ell+2$, we only take the dominant contribution into account when scaling the particle partial width with energy (see Eq. (16) of Paper I); (ii) when integrating the rate contribution of a capture resonance numerically, and the $\gamma$-ray decay strength is fragmented, we only take the strongest primary transition (and the associated final state excitation energy) into account when scaling the $\gamma$-ray partial width with energy (see Eq. (17) of Paper I); (iii) when upper limits of partial widths are involved, the code samples over a Porter-Thomas distribution of dimensionless reduced widths, with a mean value of $\langle \theta^2_i \rangle$. The mean value is found from a least-squares fit, as discussed in Sec. 5.2.1 of Paper I, but the {\it uncertainty of the mean value} is not taken into account in the present version of the code; (iv) in some cases (i.e., for proton-induced reactions on $^{18}$F, $^{19}$Ne, $^{23}$Mg, $^{24}$Al, $^{29}$P, $^{32}$Cl and $^{39}$Ca), where information had to be extracted from the mirror nucleus, the analog assignments are not unambiguous. In principle, one could sample over a discrete distribution representing the different choices of analog assignments, but the present version of the code disregards this option; (v) for pairs of competing reactions, such as (p,$\gamma$) and (p,$\alpha$), or ($\alpha$,$\gamma$) and ($\alpha$,n), the partial widths entering in the rate calculations are correlated. Therefore, if our reaction rates are used in Monte Carlo nucleosynthesis studies, or are employed to derive the branching ratio (for example, 
$N_A\left<\sigma v\right>_{p\alpha}/N_A\left<\sigma v\right>_{p\gamma}$), then the resulting uncertainties on element abundances or branching ratios are likely overestimated. We did not implement the above options in the code since it is our intention to keep the formalism as simple and transparent as possible. Although we doubt that these issues will have major consequences, we may consider them in a future version of \texttt{RatesMC}.

The present Monte Carlo reaction rates also refine the motivation for future measurements in nuclear astrophysics. For example, if an experiment reduces the upper limit of a spectroscopic factor by an order of magnitude, then the ``upper limit" of the partial {\it classical} reaction rate is also reduced by that same factor. However, the Monte Carlo reaction rates, which are more reliable, behave in an entirely different manner, as explained at length in Sec. 4.4 of Paper II. Furthermore, there will be now a new emphasis on precise measurements of resonance energies since the associated uncertainties often give rise to long tails in the reaction rate probability density function (Secs. 4.2 and 4.3 of Paper II). 

Finally we would like to comment on ``direct" versus ``indirect" measurements. The expression ``direct" refers to the measurement of a reaction cross section of astrophysical interest. An ``indirect" measurement, on the other hand, refers to a study of some nuclear quantity (by using a reaction that is not necessarily the same as the one that occurs in the stellar plasma) from which the cross section of astrophysical interest may be partially inferred. For example, measurements of the reactions $^{21}$Na(p,p)$^{21}$Na, $^{24}$Mg(p,t)$^{22}$Mg or $^{24}$Mg($^{3}$He,n$\gamma$)$^{22}$Mg in order to study astrophysically important $^{22}$Mg states represent indirect studies of the $^{21}$Na(p,$\gamma$)$^{22}$Mg reaction, whereas an experiment using a radioactive $^{21}$Na beam on a hydrogen target is called a direct measurement of the $^{21}$Na(p,$\gamma$)$^{22}$Mg reaction\footnote{The reader should not confuse the expression ``direct measurement" with the unrelated term ``direct reaction" that denotes a single-step nuclear reaction.}. The calculation of reaction rates using results from indirect measurements necessarily involves the application of some nuclear model. Unfortunately, the systematic uncertainties introduced by a model are frequently difficult to quantify. For this reason it should be obvious that {\it a direct measurement is generally preferred over an indirect one, even if the estimated reaction rate uncertainties calculated using results from indirect measurements are relatively small.}

\section{Summary}
In the present paper of the series (Paper III) we publish the nuclear physics input data used to compute the Monte Carlo reaction rates presented in  Paper II. The reaction rates are calculated using the new method discussed in Paper I. Our input data, listed in the appendix, are based on the most reliable information that we are able to extract from the published literature. By making our recommended input data available to the community we intend to improve the reproducibility and transparency of the evaluated reaction rates. The reaction rate uncertainties given in Paper II are statistical in nature and do not account for unknown systematic errors in the nuclear input data listed here. The survey of literature for this review was concluded in November 2009. 

\section{Acknowledgement}
We would like to thank Ryan Fitzgerald for helpful comments. This work was supported in part by the U.S. Department of Energy under Contract No. DE-FG02-97ER41041. 

\clearpage
\appendix

\section{Explanation of tables} \label{sec:inptfiles}
As an example, we provide below a table with the nuclear data input used to calculate the reaction rates for a hypothetical reaction X(p,$\alpha$)Y with the code \texttt{RatesMC}. None of the entries in this table represent physical values, but are listed here for illustrative purposes only. All kinematic quantities are given in the center of mass reference frame. A row is disregarded as input if it begins with the symbol `!'. The meaning of each input row will be briefly explained. \\
\\
Sample nuclear data input:\\
\footnotesize
\begin{verbatim}
01  17X(p,a)14Y
02  ****************************************************************************************************************
03  1           ! Zproj
04  8           ! Ztarget
05  2           ! Zexitparticle (=0 when only 2 channels open)
06  1.0078      ! Aproj		
07  16.999      ! Atarget
08  4.0026      ! Aexitparticle (=0 when only 2 channels open)
09  0.5         ! Jproj
10  1.5         ! Jtarget
11  0.0	        ! Jexitparticle (=0 when only 2 channels open)
12  5261.3      ! projectile separation energy (keV)
13  4015.3      ! exit particle separation energy (=0 when only 2 channels open)
14  1.25        ! Radius parameter R0 (fm)
15  3           ! Gamma-ray channel number (=2 if ejectile is a g-ray; =3 otherwise)
16  ****************************************************************************************************************
17  1.0         ! Minimum energy for numerical integration (keV)
18  5000        ! Number of random samples (>5000 for better statistics)
19  0           ! =0 for rate output at all temperatures; =NT for rate output at selected temperatures
20  ****************************************************************************************************************
21  Nonresonant Contribution
22  S(keVb)   S'(b)      S''(b/keV)   fracErr    Cutoff Energy (keV)
23  3.1e2     -2.1e-1    4.5e-6	      0.4        1200.0
24  0.0	      0.0        0.0          0.0        0.0
25  ****************************************************************************************************************
26  Resonant Contribution
27  Note: G1 = entrance channel, G2 = exit channel, G3 = spectator channel !! Ecm, Exf in (keV); wg, Gx in (eV) !!
28  Note: if Er<0, theta^2=C2S*theta_sp^2 must be entered instead of entrance channel partial width
29  Ecm	    DEcm   wg      Dwg     Jr   G1      DG1     L1    G2     DG2   L2   G3      DG3     L3   Exf   Int
30    -5.2  0.6    0       0       1    0.022   0.010   2     30.8   2.6   0    0.77    0.21    1    0.0   1
31   477.3  1.1    0       0       4    140     32      1     110    12    3    0.13    0.02    1    0.0   1
32   510.5  2.0    6.0e-1  1.8e-1  0    0       0       0     0      0     0    0       0       0    0.0   0
33  ****************************************************************************************************************
34  Upper Limits of Resonances
35  Note: enter partial width upper limit by chosing non-zero value for PT, where PT=<theta^2> for particles and...
36  Note: ...PT=<B> for g-rays [enter: "upper_limit 0.0"]; for each resonance: # upper limits < # open channels!  
37  Ecm	   DEcm	 Jr    G1       DG1     L1   PT      G2     DG2    L2   PT      G3     DG3    L3  PT    Exf   Int
38  -3.4   0.7    1    7.9e-3   0.0     1    0.0045  15.0   2.9    1    0.0     0.33   0.07   1   0.0   0.0   1  
39  73.5   0.4    1    4.5e-7   1.8e-7  1    0.0     150.0  0.0    1    0.010   0.21   0.02   1   0.0   0.0   1  
40  ****************************************************************************************************************
41  Interference between Resonances [numerical integration only]
42  Note: + for positive, - for negative interference; +- if interference sign is unknown
43  Ecm	   DEcm   Jr   G1       DG1     L1   PT      G2      DG2   L2  PT    G3     DG3    L3  PT    Exf  
44  +- 
45  -2.3   0.6    1    6.2e-3   0.0     1    0.0045  17.0    1.1   1   0.0   0.71   0.07   1   0.0   0.0    
46  89.0   0.8    1    2.9e-9   0.8e-9  1    0.0     230.0   9.0   1   0.0   0.66   0.03   1   0.0   0.0  
47  ****************************************************************************************************************
48  Comments:
49  1. Narrow resonance information is adopted from Peterson et al. 1973.
\end{verbatim}
\normalsize
\vspace{6mm}
Explanation of input:
\vspace{4mm}
\begin{tabbing}
AAAAAAAAA\= \kill
Row 01:				\> Reaction label. \\
Row 02: 				\> Separator. \\
Row 03:				\> Projectile charge. \\
Row 04:				\> Target charge. \\
Row 05:				\> Charge of {\it exit particle}; it refers to the channel other than the entrance \\
                                         \> and the $\gamma$-ray channel; =0 if only two channels are open and radiative \\
					\> capture is the only possible reaction. \\
Row 06:				\> Projectile mass in atomic mass units. \\
Row 07:				\> Target mass in atomic mass units. \\
Row 08:				\> Mass of exit particle in atomic mass units. \\
Row 09:				\> Projectile spin. \\
Row 10:                            \> Target spin. \\
Row 11:                            \> Spin of exit particle. \\			
Row 12:			         \> Separation energy of incident particle in keV. \\
Row 13:                            \> Separation energy of exit particle in keV. \\
Row 14:                            \> Radius parameter in fm, used for calculating penetration factor. \\
Row 15:                            \> Label of $\gamma$-ray channel; channel 1 refers to the incident particle, channel \\
        				         \> 2 to the emitted quantum, and channel 3 to the spectator quantum. \\
Row 16:                            \> Separator. \\				
Row 17:                            \> Minimum energy cutoff (in keV) for numerical integration of rates. \\
Row 18:                            \> Number of random samples. \\				
Row 19:                            \> Flag for temperature output; =0 outputs results at all temperatures. \\
Row 20:                             \> Separator. \\
Rows 21-22:                       \> Comments. \\				
Row 23:                            \> Input for nonresonant contribution; \texttt{S,S',S''} are the parameters $S(0)$,\\
                                         \>  $S^{\prime}(0)$, $S^{\prime\prime}(0)$ of the astrophysical S-factor (Eq. 5 of Paper I) in units  \\
                                         \> of keVb, b, b/keV, respectively; \texttt{fracErr} is the fractional uncertainty, \\
                                         \> $\sqrt{V[x]}/E[x]$, of the effective S-factor (Eq. 8 of Paper I); \texttt{Cutoff Energy}  \\
                                         \> labels the energy $E_{\mathrm{cutoff}}$ (in keV) at which the S-factor is cut off at   \\
                                         \> higher energies; it is related to the cutoff temperature (see Eq. 9 of  \\
                                         \> Paper I) by the expression \\
\end{tabbing}
\vspace{3mm}
\begin{equation}
T_{9,\mathrm{cutoff}} = 19.92~E_{\mathrm{cutoff}}^{3/2}/\sqrt{Z_0^2Z_1^2\frac{M_0M_1}{M_0+M_1}}\notag
\end{equation}
\vspace{3mm}
\begin{tabbing}
AAAAAAAAA\= \kill                                      
                                         \> where $E_{\mathrm{cutoff}}$ is in MeV and all other quantities have the same meaning \\
                                         \>  as in Sec. 2 of Paper I. \\
Row 24:                            \> Input for a second nonresonant contribution, if needed. \\
Row 25:                            \> Separator. \\
Rows 26-29:                       \> Comments. \\
Row 30:                            \> Input for resonance contribution, one input row for each resonance; \\
                                         \> \texttt{Ecm,DEcm}: resonance energy and $1\sigma$ uncertainty; \texttt{wg,Dwg}: resonance \\
                                         \>  strength, $\omega\gamma$, and associated uncertainty; \texttt{Jr}: resonance spin; \\
                                         \> \texttt{G1,DG1,L1}: incident particle partial width, uncertainty, orbital angular  \\
                                         \> momentum quantum number; for a subthreshold resonance ($E_r<0$)  \\
                                         \> the dimensionless reduced width (Eq. 14 of Paper I) is listed instead \\
                                         \> of the entrance channel partial width; \texttt{G2,DG2,L2}: partial width,  \\
                                         \> uncertainty, angular momentum quantum number (or multipolarity for \\
                                         \> $\gamma$-rays) of emitted quantum; \texttt{G3,DG3,L3}: partial width, uncertainty, \\
                                         \> angular momentum quantum number (or multipolarity for $\gamma$-rays) of \\
                                         \> spectator quantum; \texttt{Exf}: excitation energy of level in residual nucleus \\
                                         \>  that is populated in primary transition; \texttt{Int}: =0 for analytical rate  \\
                                         \> calculation; =1 for numerical rate calculation (see Eq. 1 of Paper I); for  \\
                                         \> $E_r<0$ the rate contribution is always computed numerically; when the \\
                                         \> resonance strength is entered the rate contribution is always computed \\
                                         \> analytically, regardless of the flag value; \texttt{Ecm,DEcm,Exf} are in units of \\
                                         \> keV, while resonance strengths and  partial widths are in eV. \\
Row 31:                            \> Input for second resonance; in this example, the rate contribution is \\
                                         \> calculated from partial widths and the rate is chosen to be integrated \\
                                         \> numerically. \\
Row 32:                            \> Input for third resonance; in this example, the rate contribution is \\
                                         \> calculated from the resonance strength and the rate is necessarily \\
                                         \> computed analytically. \\
Row 33:                            \> Separator. \\
Rows 34-37:                      \> Comments. \\                                      
Row 38:                            \> Input for resonance contribution when only a partial (or reduced) width \\
                                         \> upper limit is available for at least one reaction channel; the number of \\
                                         \> upper limit channels must be less than the number of open channels; \\
                                         \>  the meaning of the input quantities is the same as for row 30, except \\
                                         \> that (i) resonance strengths are not allowed as input, and (ii) the mean \\
                                         \> value, \texttt{PT}, for the Porter-Thomas distribution of dimensionless reduced \\
                                         \> widths is entered for each upper limit channel (Sec. 5.2.1 of Paper I); \\
                                         \> upper limit channels are identified by a non-zero value for the partial \\
                                         \> (or reduced) width and a zero value for the corresponding uncertainty; \\
                                         \> in this example, an upper limit for the dimensionless reduced proton \\
                                         \> width is entered (since $E_r<0$). \\
Row 39:                            \> Input for second resonance; in this example, an upper limit for the \\
                                         \> dimensionless reduced $\alpha$-particle width is entered. \\
Row 40:                            \> Separator. \\
Rows 41-43:                      \> Comments. \\                                      
Row 44:                            \> Flag for interference between resonances of same spin and parity (Sec.  \\
                                         \> 2.4 of Paper I); \texttt{+}: positive interference; \texttt{-}: negative interference; \texttt{+-}:  \\
                                         \> unknown interference sign (a binary probability density function is then\\
                                         \> used for the random sampling; Sec. 4.4 of Paper I). \\ 
Rows 45-46:                       \> Input for two interfering resonances; the meaning of the input quantities \\
                                         \> is the same as for row 38, except that the rate contribution is always \\
                                         \> computed numerically. \\
Row 47:                            \> Separator. \\
Rows 48-49:                       \> Comments; references given in this section are summarized again after \\
                                         \> each table. The symbols \texttt{Gp}, \texttt{Ga}, \texttt{Gn}, \texttt{Gg} and \texttt{G} refer to the proton,  \\
                                         \> $\alpha$-particle, neutron, $\gamma$-ray partial width and total width, respectively; \\
                                         \> \texttt{Er} denotes the resonance energy in the center of mass system. \\
\end{tabbing}
%

%
%
\clearpage
\section{Nuclear physics input} \label{sec:nuclinput}
\footnotesize
\begin{verbatim}
14C(p,g)15N
****************************************************************************************************************
1               ! Zproj                                                             
6               ! Ztarget                                                           
0               ! Zexitparticle (=0 when only 2 channels open)                      
1.0078          ! Aproj                                                             
14.003242       ! Atarget
1.009           ! Aexitparticle (=0 when only 2 channels open)                      
0.5             ! Jproj                                                             
0.0             ! Jtarget                                                           
0.5             ! Jexitparticle (=0 when only 2 channels open)                      
10207.42        ! projectile separation energy (keV)                                
10833.30        ! exit particle separation energy (=0 when only 2 channels open)    
1.25            ! Radius parameter R0 (fm)                                          
2               ! Gamma-ray channel number (=2 if ejectile is a g-ray; =3 otherwise)
****************************************************************************************************************
1.0             ! Minimum energy for numerical integration (keV)
5000            ! Number of random samples (>5000 for better statistics)
0               ! =0 for rate output at all temperatures; =NT for rate output at selected temperatures
****************************************************************************************************************
Non-Resonant Contribution
S(keVb)         S'(b)           S''(b/keV)      fracErr         Cutoff Energy (keV)
5.24            -1.22e-3        5.9e-7          0.4             1000.0
0.0             0.0             0.0             0.0             0.0
****************************************************************************************************************
Resonant Contribution
Note: G1 = entrance channel, G2 = exit channel, G3 = spectator channel !! Ecm, Exf in (keV); wg, Gx in (eV) !!
Note: if Er<0, theta^2=C2S*theta_sp^2 must be entered instead of entrance channel partial width
Ecm      DEcm   wg      Dwg     J      G1       DG1     L1    G2     DG2   L2   G3      DG3    L3    Exf     Int
 242.3   0.3    2.9e-4  0.5e-4  2.5    0        0       0     0      0     0    0       0       0    0.0     0
 325.9   0.5    3.7e-2  0.6e-2  2.5    0        0       0     0      0     0    0       0       0    0.0     0
 242.3   0.3    3.1e-3  0.5e-3  4.5    0        0       0     0      0     0    0       0       0    0.0     0
 494.5   0.3    0.84    0.13    2.5    0        0       0     0      0     0    0       0       0    0.0     0
 596.6   2.0    0.27    0.04    1.5    0        0       0     0      0     0    0       0       0    0.0     0
1085.4   0.7    0       0       0.5    7.7e3    2.0e3   1     0.29   0.10  1    7.0e2   2.0e2   1    0.0     1
1230.2   0.7    0       0       0.5    6.8e3    0.5e3   0     4.2    0.7   1    34.6e3  0.9e3   0    0.0     1
1408.0   4.0    0       0       0.5    400.9e3  6.3e3   0     21.2   0.8   1    4.0e3   0.2e3   0    0.0     1
2315.0   8.0    0       0       2.5    58.4e3   10.e3   2     0.34   0.15  2    0.0     0.0     2    5300.0  1
3183.0   10.0   0       0       1.5    35.0e3   10.0e3  2     3.0    0.9   1    26.0e3  8.0e3   0    0.0     1
****************************************************************************************************************
Upper Limits of Resonances
Note: enter partial width upper limit by chosing non-zero value for PT, where PT=<theta^2> for particles and...
Note: ...PT=<B> for g-rays [enter: "upper_limit 0.0"]; for each resonance: # upper limits < # open channels!  
Ecm     DEcm   Jr   G1      DG1     L1   PT   G2    DG2    L2   PT   G3     DG3   L3  PT     Exf   Int
! 0.0   0.0    0.0  0.0     0.0     0    0    0.0   0.0    0    0    0.0    0.0   0   0      0.0   0  
****************************************************************************************************************
Interference between Resonances [numerical integration only]
Note: + for positive, - for negative interference; +- if interference sign is unknown
Ecm	    DEcm	   Jr    G1    DG1   L1   PT    G2    DG2    L2  PT   G3    DG3    L3  PT   Exf  
!+- 
0.0     0.0     0.0   0.0   0.0   0    0     0.0   0.0    0   0    0.0   0.0    0   0    0.0  
0.0     0.0     0.0   0.0   0.0   0    0     0.0   0.0    0   0    0.0   0.0    0   0    0.0  
****************************************************************************************************************
Comments:
1. Resonance energies are derived from excitation energies (Ajzenberg-Selove 1991) and Q-value (Audi et al. 
   2003).
2. Resonance strengths are taken from Goerres et al. 1990 for the five low energy resonances.
3. Partial widths for the next five resonances are found in Bartholomew et al. 1955, Ferguson and Gove 1959,
   French et al. 1961, Kuan et al. (1971; 1976), Ramirez et al. 1972, and Niecke et al. 1977.
4. As in Goerres et al. 1990, direct capture is calculated using spectroscopic factors from Bommer et al. 1975.
\end{verbatim}
\normalsize
\vspace{5mm}
References: Ajzenberg-Selove \cite{Ajz91}; Audi et al. \cite{Aud03}; Bartholomew et al. \cite{BA55}; Bommer et al. \cite{BO75}; Ferguson and Gove \cite{FE59}; French et al. \cite{FR61}; G\"orres et al. \cite{GO90}; Kuan et al. \cite{KU71,KU76}; Niecke et al. \cite{NI77}; Ramirez et al. \cite{RA72}.
\clearpage
\footnotesize
\begin{verbatim}
14C(a,g)18O
****************************************************************************************************************
2               ! Zproj
6               ! Ztarget
0               ! Zexitparticle (=0 when only 2 channels open)
4.0026          ! Aproj         
14.003242       ! Atarget
0               ! Aexitparticle (=0 when only 2 channels open)
0.0             ! Jproj
0.0             ! Jtarget
0               ! Jexitparticle (=0 when only 2 channels open)
6226.3          ! projectile separation energy (keV)
0               ! exit particle separation energy (=0 when only 2 channels open)
1.25            ! Radius parameter R0 (fm)
2               ! Gamma-ray channel number (=2 if ejectile is a g-ray; =3 otherwise)
****************************************************************************************************************
1.0             ! Minimum energy for numerical integration (keV)
5000            ! Number of random samples (>5000 for better statistics)
0               ! =0 for rate output at all temperatures; =NT for rate output at selected temperatures
****************************************************************************************************************
Nonresonant Contribution
S(keVb)  S'(b)   S''(b/keV)   fracErr   Cutoff Energy (keV)
3.98e3   1.6     1.64e-3      1.53      2000.0
0.0      0.0     0.0          0.0       0.0
****************************************************************************************************************
Resonant Contribution
Note: G1 = entrance channel, G2 = exit channel, G3 = spectator channel !! Ecm, Exf in (keV); wg, Gx in (eV) !!
Note: if Er<0, theta^2=C2S*theta_sp^2 must be entered instead of entrance channel partial width
Ecm      DEcm   wg      Dwg    J      G1       DG1      L1   G2      DG2     L2   G3   DG3   L3   Exf     Int
 -28.1   0.7    0       0      1.0    0.025    0.020    1    0.178   0.029   1    0    0     0    0.0     1
 890.6   1.3    0.45    0.09   3.0    0        0        0    0       0       0    0    0     0    0.0     0
1389.6   0.9    1.23    0.25   4.0    0        0        0    0       0       0    0    0     0    0.0     0
1638.0   5.0    0.47    0.14   1.0    0        0        0    0       0       0    0    0     0    0.0     0
1811.5   0.9    3.22    0.64   1.0    0        0        0    0       0       0    0    0     0    0.0     0
1899.0   2.0    2.92    0.58   5.0    0        0        0    0       0       0    0    0     0    0.0     0
1987.0   4.0    0       0      2.0    1.0e3    0.8e3    2    0.412   0.093   2    0    0     0    1982.0  1
2056.0   3.0    0       0      3.0    8.0e3    1.0e3    3    0.488   0.129   1    0    0     0    3555.0  1
***************************************************************************************************************
Upper Limits of Resonances
Note: enter partial width upper limit by chosing non-zero value for PT, where PT=<theta^2> for particles and...
Note: ...PT=<B> for g-rays [enter: "upper_limit 0.0"]; for each resonance: # upper limits < # open channels!   
Ecm    DEcm    Jr   G1       DG1   L1   PT     G2     DG2    L2   PT   G3    DG3    L3   PT   Exf      Int
178.1  1.3     3    0.9e-15  0.0   3    0.010  0.022  0.017  1    0    0.0   0.0    0    0    1980.0   1
****************************************************************************************************************
Interference between Resonances [numerical integration only]
Note: + for positive, - for negative interference; +- if interference sign is unknown
Ecm    DEcm   Jr    G1      DG1   L1   PT      G2     DG2     L2  PT   G3    DG3    L3  PT   Exf  
!+- 
0.0    0.0    0.0   0.0     0.0   0    0       0.0    0.0     0   0    0.0   0.0    0   0    0.0  
0.0    0.0    0.0   0.0     0.0   0    0       0.0    0.0     0   0    0.0   0.0    0   0    0.0  
****************************************************************************************************************
Comments:
1. Following Gai et al. 1987, Goerres et al. 1992 and Lugaro et al. 2004, we consider 9 natural parity levels, 
   including one below threshold.
2. Resonance energies are derived from excitation energies (Tilley et al. 1995) and Q-value (Audi et al. 2003).
3. Seven resonance strengths and radiative widths are taken from Gai et al. 1987.
4. Alpha-particle widths for the -28 and 178 keV resonances are extracted from the alpha transfer experiment of
   Cunsolo et al. 1981, assuming a factor of two uncertainty.
5. Non-resonant contribution is adopted from Goerres et al. 1992.
\end{verbatim}
\normalsize
\vspace{5mm}
References: Audi et al. \cite{Aud03}; Cunsolo et al. \cite{CU81}; Gai et al. \cite{GA87}; G\"orres et al. \cite{GO92}; Lugaro et al. \cite{LU04}; Tilley et al. \cite{Til95}.
\clearpage
\footnotesize
\begin{verbatim}
14N(a,g)18F
****************************************************************************************************************
2         ! Zproj                                                             
7         ! Ztarget                                                           
0         ! Zexitparticle (=0 when only 2 channels open)                     
4.0026    ! Aproj                                                             
14.003    ! Atarget                                                           
0         ! Aexitparticle (=0 when only 2 channels open)                      
0.0       ! Jproj                                                             
1.0       ! Jtarget                                                           
0         ! Jexitparticle (=0 when only 2 channels open)                      
4414.6    ! projectile separation energy (keV)                                
0         ! exit particle separation energy (=0 when only 2 channels open)    
1.25      ! Radius parameter R0 (fm)                                          
2         ! Gamma-ray channel number (=2 if ejectile is a g-ray; =3 otherwise)
****************************************************************************************************************
2.0       ! Minimum energy for numerical integration (keV)
5000      ! Number of random samples (>5000 for better statistics)
0         ! =0 for rate output at all temperatures; =NT for rate output at selected temperatures
****************************************************************************************************************
Nonresonant Contribution
S(keVb)        S'(b)        S''(b/keV)    fracErr        Cutoff Energy (keV)
8.5e1          0.0          0.0           0.79           2000.0
0.0            0.0          0.0           0.0            0.0
****************************************************************************************************************
Resonant Contribution
Note: G1 = entrance channel, G2 = exit channel, G3 = spectator channel !! Ecm, Exf in (keV); wg, Gx in (eV) !!
Note: if Er<0, theta^2=C2S*theta_sp^2 must be entered instead of entrance channel partial width
 Ecm      DEcm    wg       Dwg     Jr    G1      DG1      L1   G2     DG2    L2   G3    DG3  L3   Exf      Int
 445.4    2.1     0        0       1.0   4.5e-5  0.4e-5   1    0.011  0.003  1    0     0    0    1041.6   1 
 883.0    1.6     21.0e-3  1.5e-3  0     0       0        0    0      0      0    0     0    0    0.0      0
1087.4    2.1     0.0057   0.0010  0     0       0        0    0      0      0    0     0    0    0.0      0
1190.26   0.57    1.35     0.10    0     0       0        0    0      0      0    0     0    0    0.0      0
1257.0    0.5     0.45     0.02    0     0       0        0    0      0      0    0     0    0    0.0      0
1375.2    0.6     0.016    0.006   0     0       0        0    0      0      0    0     0    0    0.0      0
1681.8    1.2     0.066    0.013   0     0       0        0    0      0      0    0     0    0    0.0      0
1693.4    3.0     0.027    0.010   0     0       0        0    0      0      0    0     0    0    0.0      0
1825.8    0.9     1.18     0.22    0     0       0        0    0      0      0    0     0    0    0.0      0
1827.4    3.0     1.32     0.24    0     0       0        0    0      0      0    0     0    0    0.0      0
1895.9    0.9     0.17     0.04    0     0       0        0    0      0      0    0     0    0    0.0      0
1970.9    1.8     0.53     0.13    0     0       0        0    0      0      0    0     0    0    0.0      0
2070.3    1.6     0.053    0.020   0     0       0        0    0      0      0    0     0    0    0.0      0
2152.4    1.6     0.097    0.020   0     0       0        0    0      0      0    0     0    0    0.0      0
2229.1    0.9     0.90     0.17    0     0       0        0    0      0      0    0     0    0    0.0      0
2362.4    1.5     0.04     0.02    0     0       0        0    0      0      0    0     0    0    0.0      0
****************************************************************************************************************
Upper Limits of Resonances
Note: enter partial width upper limit by chosing non-zero value for PT, where PT=<theta^2> for particles and...
Note: ...PT=<B> for g-rays [enter: "upper_limit 0.0"]; for each resonance: # upper limits < # open channels!  
Ecm     DEcm  Jr    G1       DG1   L1   PT       G2    DG2    L2   PT    G3    DG3    L3    PT    Exf      Int
237.4   2.1   4.0   4.1e-15  0.0   4    0.00001  0.1   0.01   1    0.0   0     0      0     0.0   1121.4   0
****************************************************************************************************************
Interference between Resonances [numerical integration only]
Note: + for positive, - for negative interference; +- if interference sign is unknown
Ecm     DEcm    Jr    G1    DG1   L1   PT    G2    DG2    L2  PT   G3    DG3    L3  PT   Exf  
!+- 
0.0     0.0     0.0   0.0   0.0   0    0     0.0   0.0    0   0    0.0   0.0    0   0    0.0  
0.0     0.0     0.0   0.0   0.0   0    0     0.0   0.0    0   0    0.0   0.0    0   0    0.0  
****************************************************************************************************************
Comments: 
1. Direct capture S-factor from Goerres et al. 2000. Note that it is not based on experimental spectroscopic
   factors, but was estimated using an alpha-particle spectroscopic factor of Sa=0.1 for all final states; we
   assign a factor 2 uncertainty to the S-factor (yielding a fractional uncertainty of 0.79; see the numerical
   example at the end of Sec. 5.1.2 of Paper I). 
2. Er=237 keV: (i) for estimation of upper limit on alpha-particle partial width, see comments in Paper II;
   (ii) value of Gg=0.1+-0.01 eV is a guess (inconsequential since Ga<<Gg).  
3. Er=445 keV: partial widths computed using measured resonance strength (Goerres et al. 2000) and measured
   lifetime (Rolfs, Berka and Azuma 1973).  
4. Ex=4964 keV (Er=549 keV with Jp;T=2+;1) disregarded (isospin-forbidden decay).
5. Ex=5603/5605 keV (Er=1189/1190 keV): doublet; measured summed strength adopted here.
6. Strengths for other resonances adopted from Becker et al. 1982; Goerres et al. 2000; Kieser et al. 1979; and
   Rolfs, Charlesworth and Azuma 1973.
7. Levels with Jp=0+ can be disregarded since their population as resonances in 14N+a is forbidden according to
   angular momentum selection rules.
8. We disregard Ex=4848 keV (Jp=5-; Er=434 keV) since this (unobserved) l=5 resonance is presumably negligible 
   compared to observed Er=445 keV (l=1) resonance.
\end{verbatim}
\normalsize
\vspace{5mm}
References: Becker et al. \cite{BE82}; G\"orres et al. \cite{Goe00}; Kieser et al. \cite{Kie79}; Rolfs, Berka and Azuma \cite{Rol73a}; Rolfs, Charlesworth and Azuma \cite{Rol73b}.
\clearpage
\footnotesize

\normalsize
\vspace{5mm}
References: Angulo et al. \cite{Ang99}; de Oliveira et al. \cite{OL96}; Wilmes et al. \cite{WI02}.
\clearpage
\footnotesize
\begin{verbatim}
15O(a,g)19Ne
****************************************************************************************************************
2               ! Zproj
8               ! Ztarget
0               ! Zexitparticle (=0 when only 2 channels open)
4.0026          ! Aproj         
15.003          ! Atarget
0               ! Aexitparticle (=0 when only 2 channels open)
0.0             ! Jproj
0.5             ! Jtarget
0               ! Jexitparticle (=0 when only 2 channels open)
3529.1          ! projectile separation energy (keV)
0               ! exit particle separation energy (=0 when only 2 channels open)
1.25            ! Radius parameter R0 (fm)
2               ! Gamma-ray channel number (=2 if ejectile is a g-ray; =3 otherwise)
****************************************************************************************************************
1.0             ! Minimum energy for numerical integration (keV)
5000            ! Number of random samples (>5000 for better statistics)
0               ! =0 for rate output at all temperatures; =NT for rate output at selected temperatures
****************************************************************************************************************
Nonresonant Contribution
S(keVb)  S'(b)   S''(b/keV)   fracErr   Cutoff Energy (keV)
2.0e4    0.0     0.0          0.4       1000.0
0.0      0.0     0.0          0.0       0.0
****************************************************************************************************************
Resonant Contribution
Note: G1 = entrance channel, G2 = exit channel, G3 = spectator channel !! Ecm, Exf in (keV); wg, Gx in (eV) !!
Note: if Er<0, theta^2=C2S*theta_sp^2 must be entered instead of entrance channel partial width
Ecm      DEcm   wg      Dwg    J     G1       DG1      L1   G2      DG2    L2   G3   DG3   L3   Exf   Int
 505.4   1.0    0       0      1.5   1.0e-5   1.5e-5   1    0.082   0.017  1    0    0     0    0.0   1
 848.7   0.8    0       0      3.5   1.5e-4   2.3e-4   3    0.458   0.092  1    0    0     0    0.0   1
1018.6   1.2    0       0      0.5   3.5e-3   0.9e-3   0    0.032   0.005  1    0    0     0    0.0   1
1072.7   1.0    0       0      2.5   0.026    0.007    3    0.060   0.016  1    0    0     0    0.0   1
1183.0   10.0   0       0      2.5   0.24     0.09     2    0.043   0.008  1    0    0     0    0.0   1   
****************************************************************************************************************
Upper Limits of Resonances
Note: enter partial width upper limit by chosing non-zero value for PT, where PT=<theta^2> for particles and...
Note: ...PT=<B> for g-rays [enter: "upper_limit 0.0"]; for each resonance: # upper limits < # open channels!   
Ecm    DEcm    Jr    G1    DG1   L1   PT   G2    DG2    L2   PT   G3    DG3    L3   PT   Exf    Int
!0.0   0.0     0.0   0.0   0.0   0    0    0.0   0.0    0    0    0.0   0.0    0    0    0.0    0
****************************************************************************************************************
Interference between Resonances [numerical integration only]
Note: + for positive, - for negative interference; +- if interference sign is unknown
Ecm    DEcm   Jr    G1      DG1   L1   PT      G2     DG2     L2  PT   G3    DG3    L3  PT   Exf  
!+- 
0.0    0.0    0.0   0.0     0.0   0    0       0.0    0.0     0   0    0.0   0.0    0   0    0.0  
0.0    0.0    0.0   0.0     0.0   0    0       0.0    0.0     0   0    0.0   0.0    0   0    0.0  
****************************************************************************************************************
Comments: 
1. Resonance energies are deduced from the excitation energies given by Tilley et al. 1995 and Tan et al. 2005.
2. Total widths are deduced from DSAM measurements by Tilley et al. 1995, Tan et al. 2005, Kanungo et al. 2006
   and Mythili et al. 2008.
3. Alpha-particle widths for the first two resonances are deduced from measurements of alpha-particle transfer 
   to the mirror levels, performed by Mao et al. 1995 and de Oliveira et al. 1996.
4. Other alpha-particle widths are obtained from measured alpha-particle branching ratios (Davids et al. 2003).        
5. A constant value of 20 MeV b is adopted for the direct capture contribution (Langanke et al. 1986; Dufour and
   Descouvemont 2000).
6. Above 0.6 GK, the rate is matched using results from the TALYS statistical model code (Goriely et al. 2008). 
\end{verbatim}
\normalsize
\vspace{5mm}
References: Davids et al. \cite{Dav03b}; Dufour and Descouvemont \cite{Duf00}; Goriely, Hilaire and Koning \cite{Gor08}; Kanungo et al. \cite{Kan06}; Langanke et al. \cite{Lan86}; Mao, Fortune and Lacaze \cite{Mao95}; Mythili et al. \cite{Myt08}; de Oliveira et al. \cite{OL96}; Tan et al. \cite{Tan05}; Tilley \cite{Til95}.   
\clearpage
\footnotesize
\begin{verbatim}
16O(a,g)20Ne
****************************************************************************************************************
2         ! Zproj                                                             
8         ! Ztarget                                                           
0         ! Zexitparticle (=0 when only 2 channels open)                     
4.0026    ! Aproj                                                             
15.9949   ! Atarget                                                           
0         ! Aexitparticle (=0 when only 2 channels open)                      
0.0       ! Jproj                                                             
0.0       ! Jtarget                                                           
0         ! Jexitparticle (=0 when only 2 channels open)                      
4729.85   ! projectile separation energy (keV)                                
0         ! exit particle separation energy (=0 when only 2 channels open)    
1.25      ! Radius parameter R0 (fm)                                          
2         ! Gamma-ray channel number (=2 if ejectile is a g-ray; =3 otherwise)
****************************************************************************************************************
2.0       ! Minimum energy for numerical integration (keV)
5000      ! Number of random samples (>5000 for better statistics)
0         ! =0 for rate output at all temperatures; =NT for rate output at selected temperatures
****************************************************************************************************************
Non-Resonant Contribution
S(keVb)        S'(b)        S''(b/keV)    fracErr        Cutoff Energy (keV)
3.62e3         -0.6872      0.0           0.4            1500.0
0.0            0.0          0.0           0.0            0.0
****************************************************************************************************************
Resonant Contribution
Note: G1 = entrance channel, G2 = exit channel, G3 = spectator channel !! Ecm, Exf in (keV); wg, Gx in (eV) !!
Note: if Er<0, theta^2=C2S*theta_sp^2 must be entered instead of entrance channel partial width
 Ecm      DEcm    wg      Dwg     Jr    G1      DG1     L1   G2      DG2     L2   G3    DG3  L3   Exf     Int
 891.6    1.7     0       0       3     2.4e-3  0.7e-3  3    2.7e-4  0.8e-4  1    0     0    0    1633.7  1
1057.9    2.6     0       0       1     28.0    3.0     1    6.7e-3  1.0e-3  1    0     0    0    1633.7  1
1995.2    5.0     0       0       0     1.9e4   0.9e3   0    7.1e-2  1.2e-2  2    0     0    0    1633.7  1
2426.5    0.5     0       0       3     8.2e3   0.3e3   3    1.5e-3  0.2e-3  1    0     0    0    4247.7  1  
2461.2    3.0     0       0       0     3.4e3   0.2e3   0    4.4e-3  0.8e-3  2    0     0    0    1633.7  1
2692.1    1.2     0       0       2     8.0e3   1.0e3   2    3.0e-2  4.0e-3  1    0     0    0    1633.7  1
3103.6    1.5     0       0       2     2.0e3   0.5e3   2    6.9e-2  0.7e-2  2    0     0    0    0.0     1
3978.2    7.0     0.21    0.05    0     0       0       0    0       0       0    0     0    0    0.0     0  
4047.8    2.2     1.35    0.15    0     0       0       0    0       0       0    0     0    0    0.0     0 
4301.2    7.0     3.05    0.38    0     0       0       0    0       0       0    0     0    0    0.0     0 
4386.2    3.0     0.18    0.02    0     0       0       0    0       0       0    0     0    0    0.0     0 
4753.2    3.0     1.3     0.5     0     0       0       0    0       0       0    0     0    0    0.0     0 
5260.2    8.0     8.0     3.0     0     0       0       0    0       0       0    0     0    0    0.0     0 
5543.4    1.9     19.5    1.5     0     0       0       0    0       0       0    0     0    0    0.0     0 
6360.2    3.0     30.2    3.5     0     0       0       0    0       0       0    0     0    0    0.0     0 
6540.2    5.0     2.06    0.25    0     0       0       0    0       0       0    0     0    0    0.0     0 
6828.2    4.0     0.41    0.05    0     0       0       0    0       0       0    0     0    0    0.0     0 
7198.2    4.0     0.23    0.05    0     0       0       0    0       0       0    0     0    0    0.0     0 
7221.2    4.0     0.131   0.002   0     0       0       0    0       0       0    0     0    0    0.0     0 
7491.2    4.0     1.41    0.23    0     0       0       0    0       0       0    0     0    0    0.0     0 
7526.2    3.0     6.6     0.8     0     0       0       0    0       0       0    0     0    0    0.0     0 
7671.2    5.0     1.94    0.15    0     0       0       0    0       0       0    0     0    0    0.0     0 
****************************************************************************************************************
Upper Limits of Resonances
Note: enter partial width upper limit by chosing non-zero value for PT, where PT=<theta^2> for particles and...
Note: ...PT=<B> for g-rays [enter: "upper_limit 0.0"]; for each resonance: # upper limits < # open channels!  
Ecm     DEcm   Jr    G1    DG1   L1   PT    G2    DG2   L2   PT   G3    DG3    L3   PT    Exf   Int
!0.0    0.0    0.0   0.0   0.0   0    0     0.0   0.0   0    0    0.0   0.0    0    0     0.0   0  
****************************************************************************************************************
Interference between Resonances [numerical integration only]
Note: + for positive, - for negative interference; +- if interference sign is unknown
Ecm     DEcm    Jr    G1    DG1   L1   PT    G2    DG2    L2  PT   G3    DG3    L3  PT   Exf  
!+- 
0.0     0.0     0.0   0.0   0.0   0    0     0.0   0.0    0   0    0.0   0.0    0   0    0.0  
0.0     0.0     0.0   0.0   0.0   0    0     0.0   0.0    0   0    0.0   0.0    0   0    0.0  
****************************************************************************************************************
Comments: 
1. Resonance strengths from Tilley et al. 1998 (see also Angulo et al. 1999).
2. For Er<3.5 MeV (except for lowest resonances; see below), G and wg from Tilley et al. 1998, Mayer 2001 and
   Angulo et al. 1999 used to calculate partial widths.
3. Direct capture S-factor estimated from transitions shown in Fig. 1 of Mohr 2005, but disregarding the 
   transitions due to Er=1058 keV resonance (they taken into account in the resonance input); we increased the
   uncertainty slightly (40%) compared to Mohr 2005 (30%).
4. Er=1058 keV: partial widths calculated from wg (Tilley et al. 1998) and Ga=28+-3 eV (MacArthur et al. 1980).
5. Er=892 keV: partial widths calculated from wg (Tilley et al. 1998) and tau=246 fs (weighted average of 
   values in Ajzenberg-Selove 1972); solution of quadratic equation gives Gx=2.4e-3 eV and Gy=2.7e-4 eV; the 
   larger value is identified with Ga, in agreement with (6Li,d) measurement (Mao et al. 1996).
\end{verbatim}
\normalsize
\vspace{5mm}
References: Ajzenberg-Selove \cite{Ajz72}; Angulo et al. \cite{Ang99}; MacArthur et al. \cite{Mac80}; Mao, Fortune and Lacaze \cite{Mao96}; Mayer \cite{May01}; Mohr \cite{Moh05}; Tilley et al. \cite{Til98}.   
\clearpage
\footnotesize

\normalsize
\vspace{5mm}
\clearpage
\footnotesize

\normalsize
\vspace{5mm}
\clearpage
\footnotesize
\begin{verbatim}
18O(p,g)19F
****************************************************************************************************************
1               ! Zproj                                                             
8               ! Ztarget                                                           
2               ! Zexitparticle (=0 when only 2 channels open)                      
1.0078          ! Aproj                                                             
17.999          ! Atarget                                                           
4.0026          ! Aexitparticle (=0 when only 2 channels open)                      
0.5             ! Jproj                                                             
0.0             ! Jtarget                                                           
0.0             ! Jexitparticle (=0 when only 2 channels open)                      
7994.8          ! projectile separation energy (keV)                                
4013.74         ! exit particle separation energy (=0 when only 2 channels open)    
1.25            ! Radius parameter R0 (fm)                                          
2               ! Gamma-ray channel number (=2 if ejectile is a g-ray; =3 otherwise)
****************************************************************************************************************
1.0             ! Minimum energy for numerical integration (keV)
5000            ! Number of random samples (>5000 for better statistics)
0               ! =0 for rate output at all temperatures; =NT for rate output at selected temperatures
****************************************************************************************************************
Nonresonant Contribution
S(keVb)         S'(b)           S''(b/keV)      fracErr         Cutoff Energy (keV)
1.57e1          0.34e-3          -2.42e-6       0.4             1000.0                 
0.0             0.0             0.0             0.0             0.0
****************************************************************************************************************
Resonant Contribution
Note: G1 = entrance channel, G2 = exit channel, G3 = spectator channel !! Ecm, Exf in (keV); wg, Gx in (eV) !!
Note: if Er<0, theta^2=C2S*theta_sp^2 must be entered instead of entrance channel partial width
Ecm     DEcm  wg      Dwg     Jr     G1      DG1     L1  G2    DG2   L2  G3      DG3     L3  Exf   Int
  19.3  0.7   0       0       2.5    2.3e-19 0.5e-19 2   2.3   1.0   1   2.5e3   1.0e3   3   0.0   1 
 142.9  0.1   0       0       0.5    1.67e-1 0.12e-1 0   0.72  0.15  1   1.23e2  0.24e2  1   0.0   1 
 204.2  1.0   5.0e-6  1.0e-6  0      0       0       0   0     0     0   0       0       0   0.0   0 
 259.5  2.6   3.7e-5  0.5e-5  0      0       0       0   0     0     0   0       0       0   0.0   0 
 315.2  1.3   0       0       2.5    1.9e-2  0.3e-2  2   0.78  0.34  1   47.0    19.0    3   0.0   1 
 588.7  1.7   1.0e-2  0.2e-2  0      0       0       0   0     0     0   0       0       0   0.0   0 
 597.1  1.2   0       0       1.5    1.4e2   0.7e2   1   0.71  0.39  1   2.0e3   0.1e3   2   0.0   1 
 798.4  1.6   0       0       0.5    24.6e3  1.4e3   0   2.5   0.4   1   20.e3   1.0e3   1   0.0   1 
 931.9  2.8   0       0       1.5    76.0    7.0     1   0.34  0.06  1   3.5e3   0.3e3   2   0.0   1 
1106.2  4.0   0.29    0.03    0      0       0       0   0     0     0   0       0       0   0.0   0 
1172.2  1.5   0       0       0.5    0.38e3  0.03e3  0   1.4   1.0   1   5.4e3   0.38e3  1   0.0   1 
1323.2  2.1   0.08    0.01    0      0       0       0   0     0     0   0       0       0   0.0   0 
1326.2  1.2   0       0       0.5    0.22e3  0.02e3  0   3.4   1.7   1   4.7e3   0.4e3   1   0.0   1 
1541.6  2.1   0.025   0.005   0      0       0       0   0     0     0   0       0       0   0.0   0 
1571.0  3.0   0.041   0.010   0      0       0       0   0     0     0   0       0       0   0.0   0 
1580.0  4.0   0.06    0.01    0      0       0       0   0     0     0   0       0       0   0.0   0 
1591.0  3.0   0.025   0.004   0      0       0       0   0     0     0   0       0       0   0.0   0 
1672.7  1.6   0       0       1.5    2.0e3   0.6e3   2   1.0   0.4   1   1.4e3   0.4e3   1   0.0   1 
1825.2  1.2   2.8     0.7     0      0       0       0   0     0     0   0       0       0   0.0   0 
1879.2  1.9   0.13    0.04    0      0       0       0   0     0     0   0       0       0   0.0   0 
1892.0  3.0   0       0       0.5    11.e3   3.0e3   0   0.36  0.20  1   18.0e3  5.4e3   1   0.0   1 
****************************************************************************************************************
Upper Limits of Resonances
Note: enter partial width upper limit by chosing non-zero value for PT, where PT=<theta^2> for particles and...
Note: ...PT=<B> for g-rays [enter: "upper_limit 0.0"]; for each resonance: # upper limits < # open channels!  
Ecm     DEcm   Jr   G1      DG1     L1   PT   G2    DG2    L2   PT   G3     DG3   L3  PT     Exf   Int
 89.0   3.0    1.5  8.0e-8  2.5e-8  1    0    0.60  0.25   1    0    3.0e3  0.0   2   0.010  0.0   1  
****************************************************************************************************************
Interference between Resonances [numerical integration only]
Note: + for positive, - for negative interference; +- if interference sign is unknown
Ecm     DEcm    Jr    G1    DG1   L1   PT    G2    DG2    L2  PT   G3    DG3    L3  PT   Exf  
!+- 
0.0     0.0     0.0   0.0   0.0   0    0     0.0   0.0    0   0    0.0   0.0    0   0    0.0  
0.0     0.0     0.0   0.0   0.0   0    0     0.0   0.0    0   0    0.0   0.0    0   0    0.0  
****************************************************************************************************************
Comments: 
1. Almost the same input data as in Angulo et al. 1999 (NACRE) are adopted here; however, we obtain the resonant
   rate contributions by numerical integration, whenever possible.
2. Proton width for Er=19 keV from Champagne and Pitt 1986 and La Cognata et al. 2008, while total and 
   radiative widths are from a private communication quoted in Wiescher et al. 1980. 
3. Proton, total and radiative widths for Er=89 keV are adopted from Lorentz-Wirzba et al. 1979, Tilley et al. 
   1995 and a private communication quoted in Wiescher et al. 1980, respectively.
4. Direct capture S-factor adopted from Wiescher et al. 1980.
5. Above T=5 GK the rate is extrapolated using the MOST Hauser-Feshbach rate.
\end{verbatim}
\normalsize
\vspace{5mm}
References: Angulo et al. \cite{Ang99}; Champagne and Pitt \cite{CH86}; La Cognata et al. \cite{La08}; Lorentz-Wirzba et al. \cite{LO79}; Tilley et al. \cite{Til95}; Wiescher et al. \cite{WI80}.
\clearpage
\footnotesize

\normalsize
\vspace{5mm}
References: Almanza et al. \cite{AL75}; Lorentz-Wirzba et al. \cite{LO79}; Mak et al. \cite{MA78}; Murillo et al. \cite{MU79}; Orihara, Rudolf and Gorodetzky \cite{OR73}; Sellin, Newson and Bilpuch \cite{SE69}; Yagi \cite{YA62}. 
\clearpage
\footnotesize
\begin{verbatim}
18O(a,g)22Ne
****************************************************************************************************************
2         ! Zproj                                                             
8         ! Ztarget                                                           
0         ! Zexitparticle (=0 when only 2 channels open)                     
4.0026    ! Aproj                                                             
17.9992   ! Atarget                                                           
1.009     ! Aexitparticle (=0 when only 2 channels open)                      
0.0       ! Jproj                                                             
0.0       ! Jtarget                                                           
0.5       ! Jexitparticle (=0 when only 2 channels open)                      
9668.1    ! projectile separation energy (keV)                                
10364.3   ! exit particle separation energy (=0 when only 2 channels open)    
1.25      ! Radius parameter R0 (fm)                                          
2         ! Gamma-ray channel number (=2 if ejectile is a g-ray; =3 otherwise)
****************************************************************************************************************
2.0       ! Minimum energy for numerical integration (keV)
5000      ! Number of random samples (>5000 for better statistics)
0         ! =0 for rate output at all temperatures; =NT for rate output at selected temperatures
****************************************************************************************************************
Non-Resonant Contribution
S(keVb)        S'(b)        S''(b/keV)    fracErr        Cutoff Energy (keV)
6.7e4          -39.6        1.4e-2        0.5            2500.0
0.0            0.0          0.0           0.0            0.0
****************************************************************************************************************
Resonant Contribution
Note: G1 = entrance channel, G2 = exit channel, G3 = spectator channel !! Ecm, Exf in (keV); wg, Gx in (eV) !!
Note: if Er<0, theta^2=C2S*theta_sp^2 must be entered instead of entrance channel partial width
 Ecm      DEcm    wg      Dwg     Jr    G1       DG1      L1   G2     DG2   L2  G3      DG3     L3  Exf     Int
  57.0    7.0     0       0       2     2.0e-41  1.6e-41  2    0.1    0.01  1   0       0       0   0.0     0
 398.0    8.0     4.8e-7  2.4e-7  0     0        0        0    0      0     0   0       0       0   0.0     0
 469.0    7.0     7.1e-7  1.7e-7  0     0        0        0    0      0     0   0       0       0   0.0     0  
 541.6    0.8     0       0       1     7.63e-5  0.63e-5  1    11.2   3.3   1   0       0       0   0.0     1
 613.5    0.8     4.9e-4  0.4e-4  0     0        0        0    0      0     0   0       0       0   0.0     0
 628.0    0.8     1.2e-3  0.1e-3  0     0        0        0    0      0     0   0       0       0   0.0     0
 947.2    4.1     4.1e-4  1.0e-4  0     0        0        0    0      0     0   0       0       0   0.0     0
1025.8    3.3     1.3e-3  0.2e-3  0     0        0        0    0      0     0   0       0       0   0.0     0
1082.2    3.3     1.0e-3  0.2e-3  0     0        0        0    0      0     0   0       0       0   0.0     0
1190.2    3.3     2.2e-3  0.3e-3  0     0        0        0    0      0     0   0       0       0   0.0     0
1254.8    4.1     6.5e-2  0.8e-2  0     0        0        0    0      0     0   0       0       0   0.0     0
1362.0    5.0     2.0e-4  0.5e-4  0     0        0        0    0      0     0   0       0       0   0.0     0
1460.1    4.1     1.8e-3  0.3e-3  0     0        0        0    0      0     0   0       0       0   0.0     0 
1526.4    3.3     7.2e-3  1.1e-3  0     0        0        0    0      0     0   0       0       0   0.0     0 
1602.5    4.1     7.0e-3  1.1e-3  0     0        0        0    0      0     0   0       0       0   0.0     0
1799.6    4.1     0.48    0.07    0     0        0        0    0      0     0   0       0       0   0.0     0
2017.9    5.0     0       0       2     500.0    170.0    2    7.7    2.3   1   4500.0  1500.0  0   1275.0  1
2083.9   10.0     0       0       1     3410.0   1023.0   1    0.29   0.10  1   7590.0  2277.0  1   0.0     1
2217.9   10.0     0       0       1     3400.0   1133.0   1    0.71   0.24  1   6600.0  2200.0  1   0.0     1
2611.9   10.0     0       0       1     6600.0   2200.0   1    4.7    1.6   1   5.9e4   2.0e4   1   0.0     1
****************************************************************************************************************
Upper Limits of Resonances
Note: enter partial width upper limit by chosing non-zero value for PT, where PT=<theta^2> for particles and...
Note: ...PT=<B> for g-rays [enter: "upper_limit 0.0"]; for each resonance: # upper limits < # open channels!  
Ecm     DEcm   Jr    G1       DG1   L1   PT    G2    DG2   L2   PT   G3    DG3    L3   PT    Exf   Int
 174.0  8.0    0     1.5e-17  0.0   0    0.01  0.1   0.01  1    0    0     0      0    0     0.0   0
****************************************************************************************************************
Interference between Resonances [numerical integration only]
Note: + for positive, - for negative interference; +- if interference sign is unknown
Ecm     DEcm    Jr    G1    DG1   L1   PT    G2    DG2    L2  PT   G3    DG3    L3  PT   Exf  
!+- 
0.0     0.0     0.0   0.0   0.0   0    0     0.0   0.0    0   0    0.0   0.0    0   0    0.0  
0.0     0.0     0.0   0.0   0.0   0    0     0.0   0.0    0   0    0.0   0.0    0   0    0.0  
****************************************************************************************************************
Comments: 
1. Direct capture S-factor from Trautvetter et al. 1978; for different results, see Buchmann et al. 1988 or
   Descouvemont 1988.
2. For all resonances above and including Er=947 keV, energies and strengths are adopted from Trautvetter et 
   al. 1978.
3. Er=398-628 keV: resonance energies from either Endt 1998 or Vogelaar et al. 1990; resonance strengths from
   Vogelaar et al. 1990 and Dababneh et al. 2003. Note that for the very weakly observed Er=398 keV resonance
   we adopt a strength uncertainty that is larger (50%) than the published value (Dababneh et al. 2003).
4. Er=542 keV: Ga calculated from measured resonance strength; Gg is equal to the total width (Berg and Wiehard
   1979).
5. Er=57 keV: we assume Jp=2+, based on the fact that la=2 partial waves describe the measured 18O(6Li,d)22Ne 
   angular distribution better than the originally suggested Jp=3- assignment (Giesen et al. 1994); for the 
   reported alpha-particle spectroscopic factor we assume an uncertainty of a factor of 2.
6. Er=174 keV: the level is very weakly populated in the 18O(6Li,d)22Ne work of Giesen et al. 1994; in fact, 
   even an la=0 transfer cannot be excluded, as can be seen by comparing the angular distributions for this
   excitation energy range (Tab. 6 in Giesen et al. 1994); thus we treat the estimated la=0 spectroscopic
   factor as an upper limit (this procedure differs from the one in the work of Giesen et al. 1994, where an
   Jp=2+ assignment was adopted).  
7. Er=1255 keV: the total width is known for this level (G=Gn=25 keV), but not enough information is available
   to deduce the gamma-ray and alpha-particle partial widths from the measured (a,g) resonance strength; thus
   we do not take the tail of this broad resonance into account.
8. Er>2 MeV: we believe that Angulo et al. 1999 (NACRE) have misinterpreted the "Gg" values listed in Graff et 
   al. 1968 and Chouraqui et al. 1970 as resonance strengths; thus the reported NACRE strengths for these high-
   energy resonances are likely erroneous. We derive alpha-particle, gamma-ray and neutron partial widths from 
   the information provided in Graff et al. 1968, Chouraqui et al. 1970 and Goldberg et al. 2004.
\end{verbatim}
\normalsize
\vspace{5mm}
References: Berg and Wiehard \cite{Ber79}; Buchmann, D'Auria and McCorquodale \cite{Buc88}; Chouraqui et al. \cite{Cho70}; Dababneh et al. \cite{Dab03}; Descouvemont \cite{Des88}; Endt \cite{End98}; Giesen et al. \cite{Gie94}; Goldberg et al. \cite{Gol04}; Graff et al. \cite{Gra68}; Trautvetter et al. \cite{Tra78}; Vogelaar et al. \cite{VO90}. 
\clearpage
\footnotesize

\normalsize
\vspace{5mm}
\clearpage
\footnotesize
\begin{verbatim}
18F(p,g)19Ne
****************************************************************************************************************
1         ! Zproj                                                             
9         ! Ztarget                                                           
2         ! Zexitparticle (=0 when only 2 channels open)                     
1.0078    ! Aproj                                                             
18.001    ! Atarget                                                           
4.0026    ! Aexitparticle (=0 when only 2 channels open)                      
0.5       ! Jproj                                                             
1.0       ! Jtarget                                                           
0.0       ! Jexitparticle (=0 when only 2 channels open)                      
6411.2    ! projectile separation energy (keV)                                
3529.1    ! exit particle separation energy (=0 when only 2 channels open)    
1.25      ! Radius parameter R0 (fm)                                          
2         ! Gamma-ray channel number (=2 if ejectile is a g-ray; =3 otherwise)
****************************************************************************************************************
1.0       ! Minimum energy for numerical integration (keV)
5000      ! Number of random samples (>5000 for better statistics)
0         ! =0 for rate output at all temperatures; =NT for rate output at selected temperatures
****************************************************************************************************************
Nonresonant Contribution
S(keVb)        S'(b)        S''(b/keV)    fracErr        Cutoff Energy (keV)
34.0           4.1e-4       0.0           14.1           1000.0
0.0            0.0          0.0           0.0            0.0
****************************************************************************************************************
Resonant Contribution
Note: G1 = entrance channel, G2 = exit channel, G3 = spectator channel !! Ecm, Exf in (keV); wg, Gx in (eV) !!
Note: if Er<0, theta^2=C2S*theta_sp^2 must be entered instead of entrance channel partial width
Ecm       DEcm     wg    Dwg   Jr    G1       DG1      L1   G2     DG2    L2   G3      DG3    L3     Exf   Int
-121.0     6.0     0     0     0.5   0.087    0.026    0    1.3    2.1    1    1.2e4   0.4e4   1     0.0   1
   8.0     6.0     0     0     1.5   7.2e-36  2.1e-36  1    1.3    2.1    1    4.e3    2.e3    2     0.0   1 
 330.0     6.0     0     0     1.5   2.22     0.69     1    5.0    2.6    2    5.7e3   0.7e3   2   275.0   1
****************************************************************************************************************
Upper Limits of Resonances
Note: enter partial width upper limit by chosing non-zero value for PT, where PT=<theta^2> for particles and...
Note: ...PT=<B> for g-rays [enter: "upper_limit 0.0"]; for each resonance: # upper limits < # open channels!  
Ecm      DEcm   Jr    G1       DG1   L1  PT      G2    DG2    L2   PT   G3      DG3     L3   PT        Exf   Int
  26.0   9.0    0.5   2.2e-17  0.0   1   0.0045  1.3   2.1    1    0    2.16e5  0.19e5  0    0         0.0   1
 287.0   6.0    2.5   2.4e-2   0.0   2   0.0045  0.29  0.15   1    0    1.2e3   0.3e3   3    0      1616.0   1
 450.0   6.0    3.5   1.0      0.0   3   0.0045  2.3   1.2    1    0    1.2e3   0.3e3   4    0       238.0   1
 827.0   6.0    1.5   700.0    0.0   0   0.0045  1.3   2.1    1    0    6.0e3   5.2e3   1    0         0.0   1
 842.0  10.0    0.5   1.8e3    0.0   0   0.0045  1.3   2.1    1    0    2.3e4   2.0e4   1    0         0.0   1 
****************************************************************************************************************
Interference between Resonances [numerical integration only]
Note: + for positive, - for negative interference; +- if interference sign is unknown
Ecm     DEcm    Jr    G1        DG1       L1   PT    G2    DG2   L2  PT  G3      DG3     L3   PT    Exf  
+- 
 38.0   7.0     1.5   1.17e-11  0.35e-11  0    0     1.1   0.6   1   0   1.3e3   0.4e3   1    0   275.0  
664.7   1.6     1.5   1.52e4    0.10e4    0    0     1.3   2.1   1   0   2.38e4  0.12e4  1    0     0.0 
****************************************************************************************************************
Comments: 
1. Radiative widths are scaled from analog levels (Nessaraja et al. 2007) when possible. Otherwise, typical 
   values and uncertainties are adopted from statistics of radiative widths in 19F.
2. The direct capture contribution is adopted from the calculation of Utku et al. 1998 using 18O+p spectroscopic
   factors.
3. For an interpretation of the large fractional uncertainty of the direct capture S-factor, see the numerical
   example at the end of Sec. 5.1.2 in Paper I; the values of E[x]=34.0 keVb and fracErr=sqrt(V[x])/E[x]=14.1
   entered here correspond to a median value of 2.4 keVb with a factor of 10 uncertainty.
4. Otherwise, the same input data as for the 18F(p,a) reaction are used. 
\end{verbatim}
\normalsize
\vspace{5mm}
References: Nesaraja et al. \cite{NE07}; Utku et al. \cite{UT98}.
\clearpage
\footnotesize
\begin{verbatim}
18F(p,a)15O
****************************************************************************************************************
1         ! Zproj                                                             
9         ! Ztarget                                                           
2         ! Zexitparticle (=0 when only 2 channels open)                     
1.0078    ! Aproj                                                             
18.001    ! Atarget                                                           
4.0026    ! Aexitparticle (=0 when only 2 channels open)                      
0.5       ! Jproj                                                             
1.0       ! Jtarget                                                           
0.0       ! Jexitparticle (=0 when only 2 channels open)                      
6411.2    ! projectile separation energy (keV)                                
3529.1    ! exit particle separation energy (=0 when only 2 channels open)    
1.25      ! Radius parameter R0 (fm)                                          
3         ! Gamma-ray channel number (=2 if ejectile is a g-ray; =3 otherwise)
****************************************************************************************************************
1.0       ! Minimum energy for numerical integration (keV)
5000      ! Number of random samples (>5000 for better statistics)
0         ! =0 for rate output at all temperatures; =NT for rate output at selected temperatures
****************************************************************************************************************
Nonresonant Contribution
S(keVb)        S'(b)        S''(b/keV)    fracErr        Cutoff Energy (keV)
0.0            0.0          0.0           0.0            0.0
0.0            0.0          0.0           0.0            0.0
****************************************************************************************************************
Resonant Contribution
Note: G1 = entrance channel, G2 = exit channel, G3 = spectator channel !! Ecm, Exf in (keV); wg, Gx in (eV) !!
Note: if Er<0, theta^2=C2S*theta_sp^2 must be entered instead of entrance channel partial width
 Ecm      DEcm    wg      Dwg    Jr    G1       DG1      L1   G2     DG2    L2   G3    DG3  L3   Exf   Int
-121.0    6.0     0       0      0.5   0.087    0.026    0    1.2e4  0.4e4  1    1.3   2.1  1    0.0   1
   8.0    6.0     0       0      1.5   7.2e-36  2.1e-36  1    4.e3   2.e3   2    1.3   2.1  1    0.0   1 
 330.0    6.0     0       0      1.5   2.22     0.69     1    5.7e3  0.7e3  2    5.0   2.6  2    0.0   1
****************************************************************************************************************
Upper Limits of Resonances
Note: enter partial width upper limit by chosing non-zero value for PT, where PT=<theta^2> for particles and...
Note: ...PT=<B> for g-rays [enter: "upper_limit 0.0"]; for each resonance: # upper limits < # open channels!  
Ecm      DEcm  Jr   G1       DG1   L1  PT      G2      DG2     L2   PT    G3    DG3    L3    PT    Exf    Int
  26.0   9.0   0.5  2.2e-17  0.0   1   0.0045  2.16e5  0.19e5  0    0     1.3   2.1    1     0     0.0    1
 287.0   6.0   2.5  2.4e-2   0.0   2   0.0045  1.2e3   0.3e3   3    0     0.29  0.15   1     0     0.0    1
 450.0   6.0   3.5  1.0      0.0   3   0.0045  1.2e3   0.3e3   4    0     2.3   1.2    1     0     0.0    1
 827.0   6.0   1.5  700.0    0.0   0   0.0045  6.0e3   5.2e3   1    0     1.3   2.1    1     0     0.0    1
 842.0  10.0   0.5  1.8e3    0.0   0   0.0045  2.3e4   2.0e4   1    0     1.3   2.1    1     0     0.0    1 
****************************************************************************************************************
Interference between Resonances [numerical integration only]
Note: + for positive, - for negative interference; +- if interference sign is unknown
Ecm     DEcm    Jr    G1        DG1       L1   PT  G2      DG2     L2  PT   G3    DG3    L3  PT   Exf  
+- 
 38.0   7.0     1.5   1.17e-11  0.35e-11  0    0   1.3e3   0.4e3   1   0    1.1   0.6    1   0    0.0  
664.7   1.6     1.5   1.52e4    0.10e4    0    0   2.38e4  0.12e4  1   0    1.3   2.1    1   0    0.0 
****************************************************************************************************************
Comments: 
1. Summary tables from Utku et al. 1998, Bardayan et al. 2004, Chae et al. 2006 and Nessaraja et al. 2007 were 
   considered, but original data were preferred.
2. Level energies mostly come from Utku et al. 1998. 
3. Alpha-particle widths have been measured by Utku et al. 1998, Bardayan et al. 2001, 2004, or are scaled 
   (Bardayan et al. 2005) from analog levels.
4. Partial widths or resonance strengths for the 330 and 665 keV resonances have been measured directly by
   Bardayan et al. (2001, 2002).
5. Recent neutron and proton transfer experiments by Adekola 2009 have reassigned spins and parities for the 
   8 keV (1/2-, 3/2-) and 38 keV (3/2-,3/2+) resonances previously thought to both have 3/2+ (Utku et al. 
   1998), interfering with the 665 keV resonance. Here we assume 3/2- and 3/2+, respectively, for these two 
   resonances and use the corresponding proton widths extracted by Adekola 2009. We allow for interferences 
   between the 38 and 665 keV resonances. Data for the subthreshold level also originate from Adekola's thesis.     
6. Reported resonances above 900 keV are not considered because of conflicting experimental data. 
\end{verbatim}
\normalsize
\vspace{5mm}
References: Adekola \cite{AD09}; Bardayan et al. \cite{BA01}; Bardayan et al. \cite{BA02}; Bardayan et al. \cite{BA04}; Bardayan et al. \cite{BA05}; Chae et al. \cite{CH06}; Nesaraja et al. \cite{NE07}; Utku et al. \cite{UT98}.
\clearpage
\footnotesize
\begin{verbatim}
19Ne(p,g)20Na
****************************************************************************************************************
1         ! Zproj                                                             
10        ! Ztarget                                                           
0         ! Zexitparticle (=0 when only 2 channels open)                      
1.0078    ! Aproj                                                             
19.002    ! Atarget                                                           
0         ! Aexitparticle (=0 when only 2 channels open)                      
0.5       ! Jproj                                                             
0.5       ! Jtarget                                                           
0.0       ! Jexitparticle (=0 when only 2 channels open)                      
2193.0    ! projectile separation energy (keV)                                
0.0       ! exit particle separation energy (=0 when only 2 channels open)    
1.25      ! Radius parameter R0 (fm)                                          
2         ! Gamma-ray channel number (=2 if ejectile is a g-ray; =3 otherwise)
****************************************************************************************************************
1.0       ! Minimum energy for numerical integration (keV)
5000      ! Number of random samples (>5000 for better statistics)
0         ! =0 for rate output at all temperatures; =NT for rate output at selected temperatures
****************************************************************************************************************
Nonresonant Contribution
S(keVb)       S'(b)        S''(b/keV)   fracErr    Cutoff Energy (keV)
1.0025        0.2288e-3    0.1376e-6    0.4        1000.0
0.0           0.0          0.0          0.0        0.0
****************************************************************************************************************
Resonant Contribution
Note: G1 = entrance channel, G2 = exit channel, G3 = spectator channel !! Ecm, Exf in (keV); wg, Gx in (eV) !!
Note: if Er<0, theta^2=C2S*theta_sp^2 must be entered instead of entrance channel partial width
Ecm     DEcm   wg      Dwg     Jr    G1      DG1    L1   G2     DG2    L2   G3   DG3   L3   Exf   Int
 452.0  9.0    9.0e-3  6.0e-3  0     0       0      0    0      0      0    0    0     0    0.0   0   
 656.0  9.0    8.0e-3  7.0e-3  0     0       0      0    0      0      0    0    0     0    0.0   0  
 808.0  7.0    0.0     0.0     1     1.98e4  2.0e3  0    0.047  0.023  1    0    0     0    0.0   1   
 893.0  7.0    0.0     0.0     0     3.59e4  2.0e3  0    0.107  0.053  1    0    0     0    0.0   1   
****************************************************************************************************************
Upper Limits of Resonances
Note: enter partial width upper limit by chosing non-zero value for PT, where PT=<theta^2> for particles and...
Note: ...PT=<B> for g-rays [enter: "upper_limit 0.0"]; for each resonance: # upper limits < # open channels!  
Ecm     DEcm    Jr    G1    DG1   L1   PT    G2    DG2    L2  PT   G3    DG3    L3  PT   Exf  Int
!0.0    0.0     0.0   0.0   0.0   0    0     0.0   0.0    0   0    0.0   0.0    0   0    0.0  0  
****************************************************************************************************************
Interference between Resonances [numerical integration only]
Note: + for positive, - for negative interference; +- if interference sign is unknown
Ecm     DEcm    Jr    G1    DG1   L1   PT    G2    DG2    L2  PT   G3    DG3    L3  PT   Exf  
!+- 
0.0     0.0     0.0   0.0   0.0   0    0     0.0   0.0    0   0    0.0   0.0    0   0    0.0  
0.0     0.0     0.0   0.0   0.0   0    0     0.0   0.0    0   0    0.0   0.0    0   0    0.0  
****************************************************************************************************************
Comments: 
1. Level energies from Tilley et al. 1998; proton separation energy from Audi et al. 2003.
2. Proton widths from Tilley et al. 1998 (Coszach et al. 1994).
3. Radiative width from shell model with 50% assigned uncertainty plus spin assignment uncertainty for the first 
   two levels.
4. Direct capture S-factor adopted from Vancraeynest et al. 1998; note that the energy in their Eq. (10) must be
   in units of MeV although their S-factor is in units of keVb.
\end{verbatim}
\normalsize
\vspace{5mm}
References: Audi et al. \cite{Aud03}; Coszach et al. \cite{CO94}; Tilley et al. \cite{Til98}; Vancraeynest et al. \cite{VA98}.
\clearpage
\footnotesize
\begin{verbatim}
20Ne(p,g)21Na
****************************************************************************************************************
1               ! Zproj
10              ! Ztarget
0               ! Zexitparticle (=0 when only 2 channels open)
1.008           ! Aproj         
19.992          ! Atarget
0               ! Aexitparticle (=0 when only 2 channels open)
0.5             ! Jproj
0.0             ! Jtarget
0               ! Jexitparticle (=0 when only 2 channels open)
2431.69         ! projectile separation energy (keV)
0               ! exit particle separation energy (=0 when only 2 channels open)
1.25            ! Radius parameter R0 (fm)
2               ! Gamma-ray channel number (=2 if ejectile is a g-ray; =3 otherwise)
****************************************************************************************************************
1.0             ! Minimum energy for numerical integration (keV)
5000            ! Number of random samples (>5000 for better statistics)
0               ! =0 for rate output at all temperatures; =NT for rate output at selected temperatures
****************************************************************************************************************
Nonresonant Contribution
S(keVb)  S'(b)     S''(b/keV)  fracErr    Cutoff Energy (keV)
18.14    -8.93e-3  5.776e-6    0.24       2000.0
19.68    -0.144    7.394e-4    0.45       239.0
****************************************************************************************************************
Resonant Contribution
Note: G1 = entrance channel, G2 = exit channel, G3 = spectator channel !! Ecm, Exf in (keV); wg, Gx in (eV) !!
Note: if Er<0, theta^2=C2S*theta_sp^2 must be entered instead of entrance channel partial width
Ecm      DEcm   wg      Dwg      J     G1      DG1     L1   G2     DG2    L2    G3     DG3   L3   Exf     Int
 -6.79   0.42   0       0        0.5   0.445   0.099   0    0.17   0.05   1     0      0     0    0.0     1
 366.2   0.5    1.1e-4  2.0e-5   0     0       0       0    0      0      0     0      0     0    0.0     0
 397.4   0.7    6.2e-5  1.2e-5   0     0       0       0    0      0      0     0      0     0    0.0     0
1112.6   0.4    1.125   1.8e-2   0     0       0       0    0      0      0     0      0     0    0.0     0
1247.2   0.4    0.035   0.01     0     0       0       0    0      0      0     0      0     0    0.0     0
1430.5   0.5    0.050   0.015    0     0       0       0    0      0      0     0      0     0    0.0     0
1737.9   0.7    0       0        1.5   1.8e5   1.5e4   1    0.50   0.15   1     0      0     0    332.0   1
1862.6   0.6    2.40    0.70     0     0       0       0    0      0      0     0      0     0    0.0     0
2036.2   0.7    0       0        1.5   2.1e4   3.0e3   2    0.80   0.20   1     0      0     0    0.0     1
****************************************************************************************************************
Upper Limits of Resonances
Note: enter partial width upper limit by chosing non-zero value for PT, where PT=<theta^2> for particles and...
Note: ...PT=<B> for g-rays [enter: "upper_limit 0.0"]; for each resonance: # upper limits < # open channels!   
Ecm      DEcm   Jr    G1     DG1    L1    PT     G2    DG2   L2   PT    G3    DG3   L3   PT   Exf   Int
!0.0     0.0    0.0   0.0    0.0    0     0      0.0   0.0   0    0     0.0   0.0   0    0    0.0   0
****************************************************************************************************************
Interference between Resonances [numerical integration only]
Note: + for positive, - for negative interference; +- if interference sign is unknown
Ecm    DEcm   Jr    G1      DG1   L1   PT      G2     DG2     L2  PT   G3    DG3    L3  PT   Exf
!+- 
0.0    0.0    0.0   0.0     0.0   0    0       0.0    0.0     0   0    0.0   0.0    0   0    0.0 
0.0    0.0    0.0   0.0     0.0   0    0       0.0    0.0     0   0    0.0   0.0    0   0    0.0 
****************************************************************************************************************
Comments:
1. Nonresonant contribution from fits to direct capture data of Rolfs et al. 1975, renormalized using revised 
   strength for the 1112.6 keV resonance and re-analysis of 16O(p,g) direct capture (Iliadis et al. 2008). Two 
   terms are necessary to cover the full energy range of Rolfs et al. 1975.
2. Energy of subthreshold state is based on Q=2431.69(14) keV, which includes a new measurement of the mass of 
   21Na by Mukherjee et al. 2008. Spectroscopic factor of 0.71(14) is a weighted average of fits to direct 
   capture and the resonance tail. The dimensionless reduced width is 0.609(61) (Iliadis 1997) and the gamma-ray 
   partial width is from the lifetime measurement of Anttila et al. 1977.
3. Information on resonances: adopted energies from ENSDF (Firestone 2004), widths and strengths from Rolfs et 
   al. 1975, and Endt and van der Leun 1978.
\end{verbatim}
\normalsize
\vspace{5mm}
References: Anttila et al. \cite{antilla}; Endt and van der Leun \cite{End78}; Firestone \cite{endsf}; Iliadis \cite{ili2}; Iliadis et al. \cite{Ili08}; Mukherjee et al. \cite{muk2}; Rolfs et al. \cite{Rol75}.
\clearpage
\footnotesize

\normalsize
\vspace{5mm}
References: Anantaraman et al. \cite{raman}; Endt and van der Leun \cite{End78}; Fifield et al. \cite{fif}; Firestone \cite{endsf2}; Highland and Thwaites \cite{high}; Schmalbrock et al. \cite{sch}; Smulders \cite{smu}.
\clearpage
\footnotesize

\normalsize
\vspace{5mm}
\clearpage
\footnotesize

\normalsize
\vspace{5mm}
References: Endt \cite{End90}; G\"orres et al. \cite{Gor83}; Hale et al. \cite{Hal01}; Powers et al. \cite{Pow71}.
\clearpage
\footnotesize

\normalsize
\vspace{5mm}
References: Longland et al. \cite{Lon09}.
\clearpage
\footnotesize

\normalsize
\vspace{5mm}
References: Longland et al. \cite{Lon09}.
\clearpage
\footnotesize

\normalsize
\vspace{5mm}
\clearpage
\footnotesize
\begin{verbatim}
22Na(p,g)23Mg
****************************************************************************************************************
1               ! Zproj
11              ! Ztarget
0               ! Zexitparticle (=0 when only 2 channels open)
1.0078          ! Aproj         
21.9944         ! Atarget
0               ! Aexitparticle (=0 when only 2 channels open)
0.5             ! Jproj
3.0             ! Jtarget
0.0             ! Jexitparticle (=0 when only 2 channels open)
7580.3          ! projectile separation energy
0.0             ! exit particle separation energy (=0 when only 2 channels open)
1.25            ! Radius parameter R0 (fm)
2               ! Gamma-ray channel number (=2 if ejectile is a g-ray; =3 otherwise)
****************************************************************************************************************
1.0             ! Minimum energy for numerical integration (keV)
5000            ! Number of random samples (>5000 for better statistics)
0               ! =0 for rate output at all temperatures; =NT for rate output at selected temperatures
****************************************************************************************************************
Nonresonant contribution
S(keVb) S'(b)   S''(b/keV)      fracErr Cutoff Energy (keV)
1.8e1   0.0     0.0             0.4     1500.0
0.0     0.0     0.0             0.0     0.0
****************************************************************************************************************
Resonant Contribution
Note: G1 = entrance channel, G2 = exit channel, G3 = spectator channel !! Ecm, Exf in (keV); wg, Gx in (eV) !!
Note: if Er<0, theta^2=C2S*theta_sp^2 must be entered instead of entrance channel partial width
Ecm     DEcm    wg      Dwg     Jr      G1      DG1      L1     G2      DG2      L2   G3  DG3  L3 Exf   Int
  43.1  1.7     0       0       4.5     1.0e-16 0.4e-16  2      1.6e-1  0.8e-1   1    0   0    0  0.0   1
  66.6  3.0     0       0       1.5     1.8e-12 0.72e-12 2      2.0e-2  1.0e-2   1    0   0    0  0.0   0
 204.3  1.8     1.4e-3  0.3e-3  0       0       0        0      0       0        0    0   0    0  0.0   0
 271.2  2.0     1.6e-2  0.3e-2  0       0       0        0      0       0        0    0   0    0  0.0   0
 435.0  2.2     6.8e-2  2.0e-2  0       0       0        0      0       0        0    0   0    0  0.0   0
 480.7  2.4     3.7e-2  1.2e-2  0       0       0        0      0       0        0    0   0    0  0.0   0
 579.4  2.4     2.4e-1  0.3e-1  0       0       0        0      0       0        0    0   0    0  0.0   0
 706.7  2.4     3.6e-1  0.6e-1  0       0       0        0      0       0        0    0   0    0  0.0   0
 760.7  2.4     9.5e-2  0.3e-2  0       0       0        0      0       0        0    0   0    0  0.0   0
****************************************************************************************************************
Upper Limits of Resonances
Note: enter partial width upper limit by chosing non-zero value for PT, where PT=<theta^2> for particles and...
Note: ...PT=<B> for g-rays [enter: "upper_limit 0.0"]; for each resonance: # upper limits < # open channels!  
Ecm     DEcm    Jr    G1     DG1  L1  PT     G2    DG2    L2  PT   G3    DG3  L3  PT  Exf     Int
 188.9  1.7     2.5   0.0072 0    0   0.0045 0.33  0.16   1   0    0     0    0   0   2715.0  0 
 221.7  2.4     2.5   0.0061 0    0   0.0045 0.2   0.1    1   0    0     0    0   0   0.0     0
 493.7  6.2     2.5   0.018  0    0   0.0045 0.2   0.1    1   0    0     0    0   0   0.0     0
 613.5  8.0     2.5   0.021  0    0   0.0045 0.2   0.1    1   0    0     0    0   0   0.0     0
****************************************************************************************************************
Interference between Resonances [numerical integration only]
Note: + for positive, - for negative interference; +- if interference sign is unknown
Ecm     DEcm    Jr    G1     DG1  L1   PT    G2    DG2    L2  PT   G3    DG3  L3  PT   Exf    
!+- 
0.0     0.0     0.0   0.0    0.0  0    0     0.0   0.0    0   0    0.0   0.0  0   0    0.0    
0.0     0.0     0.0   0.0    0.0  0    0     0.0   0.0    0   0    0.0   0.0  0   0    0.0    
****************************************************************************************************************
Comments:
1. Gg=2.0e-2 eV for Er=66.6 keV resonance is a guess (not important since Gp<<Gg). 
2. Gg=2.0e-1 eV for Er=222, 494 and 614 keV resonances is a guess; we assume for these undetected resonances
   that Gp<<Gg (assumption most likely inconsequential for total rates). 
3. Spin and parity of Er=494 and 614 keV resonances unknown; for upper limit contributions we assume s-waves
   (Jp=5/2+).
\end{verbatim}
\normalsize
\vspace{5mm}
\clearpage
\footnotesize

\normalsize
\vspace{5mm}
\clearpage
\footnotesize

\normalsize
\vspace{5mm}
\clearpage
\footnotesize
\begin{verbatim}
22Mg(p,g)23Al
****************************************************************************************************************
1               ! Zproj
12              ! Ztarget
0               ! Zexitparticle (=0 when only 2 channels open)
1.0078          ! Aproj         
21.999          ! Atarget
0               ! Aexitparticle (=0 when only 2 channels open)
0.5             ! Jproj
0.0             ! Jtarget
0.0             ! Jexitparticle (=0 when only 2 channels open)
122.0           ! projectile separation energy (keV)
0.0             ! exit particle separation energy (=0 when only 2 channels open)
1.25            ! Radius parameter R0 (fm)
2               ! Gamma-ray channel number (=2 if ejectile is a g-ray; =3 otherwise)
****************************************************************************************************************
1.0             ! Minimum energy for numerical integration (keV)
7000            ! Number of random samples (>5000 for better statistics)
0               ! =0 for rate output at all temperatures; =NT for rate output at selected temperatures
****************************************************************************************************************
Nonresonant contribution
S(keVb) S'(b)   S''(b/keV)      fracErr Cutoff Energy (keV)
4.4e-1  2.7e-4  -4.5e-8         0.4     4500.0
0.0     0.0     0.0             0.0     0.0
****************************************************************************************************************
Resonant Contribution
Note: G1 = entrance channel, G2 = exit channel, G3 = spectator channel !! Ecm, Exf in (keV); wg, Gx in (eV) !!
Note: if Er<0, theta^2=C2S*theta_sp^2 must be entered instead of entrance channel partial width
Ecm     DEcm    wg      Dwg     J       G1      DG1      L1    G2      DG2      L2   G3  DG3  L3 Exf    Int
 405    27      0       0       0.5     74      30       0     7.2e-7  1.4e-7   2    0   0    0  0.0    1
1651    40      0       0       1.5     912     365      2     8.3e-4  4.2e-4   1    0   0    0  0.0    1
****************************************************************************************************************
Upper Limits of Resonances
Note: enter partial width upper limit by chosing non-zero value for PT, where PT=<theta^2> for particles and...
Note: ...PT=<B> for g-rays [enter: "upper_limit 0.0"]; for each resonance: # upper limits < # open channels!  
Ecm     DEcm    Jr    G1    DG1   L1   PT    G2    DG2    L2  PT   G3    DG3    L3  PT   Exf  Int
!0.0    0.0     0.0   0.0   0.0   0    0     0.0   0.0    0   0    0.0   0.0    0   0    0.0  0  
****************************************************************************************************************
Interference between Resonances [numerical integration only]
Note: + for positive, - for negative interference; +- if interference sign is unknown
Ecm     DEcm    Jr    G1    DG1   L1   PT    G2    DG2    L2  PT   G3    DG3    L3  PT   Exf  
!+- 
0.0     0.0     0.0   0.0   0.0   0    0     0.0   0.0    0   0    0.0   0.0    0   0    0.0  
0.0     0.0     0.0   0.0   0.0   0    0     0.0   0.0    0   0    0.0   0.0    0   0    0.0  
****************************************************************************************************************
Comments:
1. S-factor for direct capture to ground state calculated with C2S=0.22 from 23Ne mirror state.
2. Several higher-energy resonances are missing between those considered here and those observed in the
   scattering study of He et al. 2007.
\end{verbatim}
\normalsize
\vspace{5mm}
References: He et al. \cite{He07}.
\clearpage
\footnotesize

\normalsize
\vspace{5mm}
References: Endt \cite{End90}; Lotay et al. \cite{Lot08}; Tomandl et al. \cite{Tom04}; Visser et al. \cite{Vis07,Vis08}.
\clearpage
\footnotesize

\normalsize
\vspace{5mm}
\clearpage
\footnotesize

\normalsize
\vspace{5mm}
References: Cseh et al. \cite{cseh}; Draayer et al. \cite{dr74}; Endt \cite{End90}; Firestone \cite{endsf2}; Lyons et al. \cite{ly69}; Maas et al. \cite{maas}; Smulders and Endt \cite{smE}; Strandberg et al. \cite{str08}; Tanabe et al. \cite{ta83}.
\clearpage
\footnotesize

\normalsize
\vspace{5mm}
References: Endt \cite{End90}; Endt and Rolfs \cite{EnR87}; Iliadis \cite{Ili89}; Iliadis et al. \cite{Ili96}; Iliadis et al. \cite{Ili01}.
\clearpage
\footnotesize

\normalsize
\vspace{5mm}
References: Endt and Rolfs \cite{EnR87}; Iliadis \cite{Ili89}.
\clearpage
\footnotesize

\normalsize
\vspace{5mm}
References: Endt and Rolfs \cite{EnR87}; Iliadis \cite{Ili89}.
\clearpage
\footnotesize

\normalsize
\vspace{5mm}
References: Champagne et al. \cite{Cha90}; Iliadis et al. \cite{Ili90}.
\clearpage
\footnotesize
\begin{verbatim}
23Al(p,g)24Si
****************************************************************************************************************
1               ! Zproj
13              ! Ztarget
0               ! Zexitparticle (=0 when only 2 channels open)
1.0078          ! Aproj		
23.0073         ! Atarget
0               ! Aexitparticle (=0 when only 2 channels open)
0.5             ! Jproj
2.5             ! Jtarget
0.0             ! Jexitparticle (=0 when only 2 channels open)
3304.0          ! projectile separation energy (keV)
0.0             ! exit particle separation energy (=0 when only 2 channels open)
1.25            ! Radius parameter R0 (fm)
2               ! Gamma-ray channel number (=2 if ejectile is a g-ray; =3 otherwise)
****************************************************************************************************************
1.0             ! Minimum energy for numerical integration (keV)
7000            ! Number of random samples (>5000 for better statistics)
0               ! =0 for rate output at all temperatures; =NT for rate output at selected temperatures
****************************************************************************************************************
Nonresonant contribution
S(keVb)	  S'(b)	      S''(b/keV)    fracErr     Cutoff Energy (keV)
4.64e0    -6.24e-4    0.0           0.5	        2500.0
0.0       0.0         0.0           0.0	        0.0
****************************************************************************************************************
Resonant Contribution
Note: G1 = entrance channel, G2 = exit channel, G3 = spectator channel !! Ecm, Exf in (keV); wg, Gx in (eV) !!
Note: if Er<0, theta^2=C2S*theta_sp^2 must be entered instead of entrance channel partial width
Ecm     DEcm    wg      Dwg     J       G1      DG1      L1    G2      DG2      L2   G3  DG3  L3 Exf    Int
 137     29     0       0       2       9.3e-6  4.6e-6   0     1.7e-2  0.8e-2   1    0   0    0  0.0    1
 746    150     0       0       4       5.8e-1  2.9e-1   2     7.2e-4  3.6e-4   2    0   0    0  0.0    1
 866    150     0       0       0       2.9e1   1.4e1    2     3.4e-4  1.7e-4   2    0   0    0  0.0    1
1166    150     0       0       3       2.2e4   1.1e4    0     8.0e-3  4.0e-3   1    0   0    0  0.0    1
****************************************************************************************************************
Upper Limits of Resonances
Note: enter partial width upper limit by chosing non-zero value for PT, where PT=<theta^2> for particles and...
Note: ...PT=<B> for g-rays [enter: "upper_limit 0.0"]; for each resonance: # upper limits < # open channels! 
Ecm     DEcm    Jr    G1    DG1   L1   PT    G2    DG2    L2  PT   G3    DG3    L3  PT   Exf  Int
!0.0    0.0     0.0   0.0   0.0   0    0     0.0   0.0    0   0    0.0   0.0    0   0    0.0  0  
****************************************************************************************************************
Interference between Resonances [numerical integration only]
Note: + for positive, - for negative interference; +- if interference sign is unknown
Ecm     DEcm    Jr    G1    DG1   L1   PT    G2    DG2    L2  PT   G3    DG3    L3  PT   Exf  
!+- 
0.0     0.0     0.0   0.0   0.0   0    0     0.0   0.0    0   0    0.0   0.0    0   0    0.0  
0.0     0.0     0.0   0.0   0.0   0    0     0.0   0.0    0   0    0.0   0.0    0   0    0.0  
****************************************************************************************************************
Comments:
1. Present direct capture S-factor is calculated using shell-model spectroscopic factors from Herndl et al. 
   1995.
2. Three higher lying resonances correspond to unobserved levels; 150 keV energy uncertainty is a rough estimate 
   only.
3. Proton and gamma-ray partial widths are based on shell-model results of Herndl et al. 1995; assumed 
   uncertainty amounts here to 50%.
\end{verbatim}
\normalsize
\vspace{5mm}
References: Herndl et al. \cite{Her95}.
\clearpage
\footnotesize
\begin{verbatim}
24Al(p,g)25Si
****************************************************************************************************************
1               ! Zproj
13              ! Ztarget
0               ! Zexitparticle (=0 when only 2 channels open)
1.0078          ! Aproj         
23.9999         ! Atarget
0               ! Aexitparticle (=0 when only 2 channels open)
0.5             ! Jproj
4.0             ! Jtarget
0.0             ! Jexitparticle (=0 when only 2 channels open)
3408.0          ! projectile separation energy (keV)
0.0             ! exit particle separation energy (=0 when only 2 channels open)
1.25            ! Radius parameter R0 (fm)
2               ! Gamma-ray channel number (=2 if ejectile is a g-ray; =3 otherwise)
****************************************************************************************************************
1.0             ! Minimum energy for numerical integration (keV)
7000            ! Number of random samples (>5000 for better statistics)
0               ! =0 for rate output at all temperatures; =NT for rate output at selected temperatures
****************************************************************************************************************
Nonresonant contribution
S(keVb) S'(b)   S''(b/keV)      fracErr Cutoff Energy (keV)
2.7e1   0.0     0.0             0.5     5000.0
0.0     0.0     0.0             0.0     0.0
****************************************************************************************************************
Resonant Contribution
Note: G1 = entrance channel, G2 = exit channel, G3 = spectator channel !! Ecm, Exf in (keV); wg, Gx in (eV) !!
Note: if Er<0, theta^2=C2S*theta_sp^2 must be entered instead of entrance channel partial width
Ecm     DEcm    wg      Dwg     J       G1      DG1      L1    G2      DG2      L2   G3  DG3  L3 Exf    Int
 190.0  100.0   0       0       2.5     5.7e-7  2.8e-7   2     4.0e-2  2.0e-2   1    0   0    0  0.0    1
 410.0   22.0   0       0       4.5     7.5e-1  3.7e-1   0     1.3e-2  0.7e-2   1    0   0    0  0.0    1
 500.0  100.0   0       0       3.5     2.38    1.19     0     1.5e-2  0.8e-2   1    0   0    0  0.0    1
 510.0  100.0   0       0       2.5     0.1     0.05     2     7.2e-2  3.6e-2   1    0   0    0  0.0    1
 730.0  100.0   0       0       2.5     5.3e-2  2.7e-2   2     1.2e-1  0.6e-1   1    0   0    0  0.0    1
****************************************************************************************************************
Upper Limits of Resonances
Note: enter partial width upper limit by chosing non-zero value for PT, where PT=<theta^2> for particles and...
Note: ...PT=<B> for g-rays [enter: "upper_limit 0.0"]; for each resonance: # upper limits < # open channels!  
Ecm     DEcm    Jr    G1    DG1   L1   PT    G2    DG2    L2  PT   G3    DG3    L3  PT   Exf  Int
!0.0    0.0     0.0   0.0   0.0   0    0     0.0   0.0    0   0    0.0   0.0    0   0    0.0  0  
****************************************************************************************************************
Interference between Resonances [numerical integration only]
Note: + for positive, - for negative interference; +- if interference sign is unknown
Ecm     DEcm    Jr    G1    DG1   L1   PT    G2    DG2    L2  PT   G3    DG3    L3  PT   Exf  
!+- 
0.0     0.0     0.0   0.0   0.0   0    0     0.0   0.0    0   0    0.0   0.0    0   0    0.0  
0.0     0.0     0.0   0.0   0.0   0    0     0.0   0.0    0   0    0.0   0.0    0   0    0.0  
****************************************************************************************************************
Comments:
1. Energy uncertainty of 100 keV for 190, 500, 510 and 730 keV resonances is a rough guess only.
\end{verbatim}
\normalsize
\vspace{5mm}
\clearpage
\footnotesize
\begin{verbatim}
25Al(p,g)26Si
****************************************************************************************************************
1               ! Zproj
13              ! Ztarget
0               ! Zexitparticle (=0 when only 2 channels open)
1.0078          ! Aproj         
24.9904         ! Atarget
0               ! Aexitparticle (=0 when only 2 channels open)
0.5             ! Jproj
2.5             ! Jtarget
0.0             ! Jexitparticle (=0 when only 2 channels open)
5513.7          ! projectile separation energy (keV)
0.0             ! exit particle separation energy (=0 when only 2 channels open)
1.25            ! Radius parameter R0 (fm)
2               ! Gamma-ray channel number (=2 if ejectile is a g-ray; =3 otherwise)
****************************************************************************************************************
1.0             ! Minimum energy for numerical integration (keV)
5000            ! Number of random samples (>5000 for better statistics)
0               ! =0 for rate output at all temperatures; =NT for rate output at selected temperatures
****************************************************************************************************************
Non-resonant contribution
S(keVb) S'(b)   S''(b/keV)      fracErr Cutoff Energy (keV)
2.7e1   0.0     0.0             0.4     1000.0
0.0     0.0     0.0             0.0     0.0
****************************************************************************************************************
Resonant Contribution
Note: G1 = entrance channel, G2 = exit channel, G3 = spectator channel !! Ecm, Exf in (keV); wg, Gx in (eV) !!
Note: if Er<0, theta^2=C2S*theta_sp^2 must be entered instead of entrance channel partial width
Ecm     DEcm    wg      Dwg     J       G1      DG1      L1    G2      DG2      L2   G3  DG3  L3 Exf      Int
 163.3  1.8     0       0       1       7.8e-9  3.1e-9   2     0.11    0.055    1    0   0    0  1808.7   1
 407.0  2.6     0       0       3       3.9     1.6      0     0.10    0.05     1    0   0    0  4350.0   1
 438.3  4.0     0       0       0       2.3e-2  0.9e-2   2     0.0046  0.0023   1    0   0    0  1808.7   1
 806.3  4.0     0       0       1       4.6     1.8      2     0.11    0.055    1    0   0    0  2938.4   1
 882.3  4.0     0       0       2       292     117      0     0.11    0.055    1    0   0    0  1808.7   1
 965.3  4.0     0       0       4       3.2     1.3      2     0.017   0.0085   1    0   0    0  4900.3   1
****************************************************************************************************************
Upper Limits of Resonances
Note: enter partial width upper limit by chosing non-zero value for PT, where PT=<theta^2> for particles and...
Note: ...PT=<B> for g-rays [enter: "upper_limit 0.0"]; for each resonance: # upper limits < # open channels!  
Ecm     DEcm    Jr    G1    DG1   L1   PT    G2    DG2    L2  PT   G3    DG3    L3  PT   Exf  Int
!0.0    0.0     0.0   0.0   0.0   0    0     0.0   0.0    0   0    0.0   0.0    0   0    0.0  0  
****************************************************************************************************************
Interference between Resonances [numerical integration only]
Note: + for positive, - for negative interference; +- if interference sign is unknown
Ecm     DEcm    Jr    G1    DG1   L1   PT    G2    DG2    L2  PT   G3    DG3    L3  PT   Exf  
!+- 
0.0     0.0     0.0   0.0   0.0   0    0     0.0   0.0    0   0    0.0   0.0    0   0    0.0  
0.0     0.0     0.0   0.0   0.0   0    0     0.0   0.0    0   0    0.0   0.0    0   0    0.0  
****************************************************************************************************************
Comments:
1. Excitation energies of first three resonances from Wrede 2009; for last three resonances from column 1 of 
   Tab. I in Parpottas et al. 2004, but we added an average value of 8 keV, by which the excitation energies
   of Parpottas et al. seem to be too small (see comments in Wrede 2009).
2. Unlike Wrede 2009, we prefer to calculate all resonance energies from excitation energies; in particular,
   we do not use the proton energy from the beta-delayed proton decay of 26P since it is less precise than the
   value derived from Ex.
3. For the last three resonances the Jp assignments are uncertain; note that the values of 2+, 2+, 0+ (see
   column 6 of Tab. I in Parpottas et al. 2004) are inconsistent both with the known level scheme of the 26Mg
   mirror and the calculated level scheme from the shell model.
4. All spectroscopic factors and gamma-ray partial widths are adopted from the shell model (Iliadis et al. 
   1996).
5. The final states for primary gamma-ray transitions are adopted from the 26Mg mirror decay (Endt 1990).
\end{verbatim}
\normalsize
\vspace{5mm}
References: Endt \cite{End90}; Iliadis et al. \cite{Ili96}; Parpottas et al. \cite{Par04}; Wrede \cite{Wre09b}.
\clearpage
\footnotesize
\begin{verbatim}
26Al^g(p,g)27Si
****************************************************************************************************************
1               ! Zproj
13              ! Ztarget
0               ! Zexitparticle (=0 when only 2 channels open)
1.0078          ! Aproj         
25.986891       ! Atarget
0               ! Aexitparticle (=0 when only 2 channels open)
0.5             ! Jproj
5.0             ! Jtarget
0.0             ! Jexitparticle (=0 when only 2 channels open)
7462.96         ! projectile separation energy
0.0             ! exit particle separation energy (=0 when only 2 channels open)
1.25            ! Radius parameter R0 (fm)
2               ! Gamma-ray channel number (=2 if ejectile is a g-ray; =3 otherwise)
****************************************************************************************************************
1.0             ! Minimum energy for numerical integration (keV)
20000           ! Number of random samples (>5000 for better statistics)
0               ! =0 for rate output at all temperatures; =NT for rate output at selected temperatures
****************************************************************************************************************
Non-resonant contribution
S(keVb) S'(b)   S''(b/keV)      fracErr Cutoff Energy (keV)
8.0e1   0.0     0.0             0.4     2000.0
0.0     0.0     0.0             0.0     0.0
****************************************************************************************************************
Resonant Contribution
Note: G1 = entrance channel, G2 = exit channel, G3 = spectator channel !! Ecm, Exf in (keV); wg, Gx in (eV) !!
Note: if Er<0, theta^2=C2S*theta_sp^2 must be entered instead of entrance channel partial width
Ecm      DEcm   wg      Dwg     Jr      G1      DG1     L1      G2      DG2     L2   G3  DG3  L3  Exf   Int
 188.6   0.4    4.7e-5  0.7e-5  0       0       0       0       0       0       0    0   0    0   0     0       
 241.3   0.3    1.0e-5  0.5e-5  0       0       0       0       0       0       0    0   0    0   0     0
 276.3   0.4    2.9e-3  0.3e-3  0       0       0       0       0       0       0    0   0    0   0     0
 368.5   0.4    6.9e-2  0.7e-2  0       0       0       0       0       0       0    0   0    0   0     0
 693.6   0.4    6.7e-2  0.9e-2  0       0       0       0       0       0       0    0   0    0   0     0
 702.1   0.2    3.3e-2  0.5e-2  0       0       0       0       0       0       0    0   0    0   0     0
 741.8   0.2    2.1e-1  0.3e-1  0       0       0       0       0       0       0    0   0    0   0     0
 763.3   0.4    4.6e-2  0.7e-2  0       0       0       0       0       0       0    0   0    0   0     0
 815.7   1.0    3.5e-3  0.6e-3  0       0       0       0       0       0       0    0   0    0   0     0
 827.1   0.4    5.2e-2  0.7e-2  0       0       0       0       0       0       0    0   0    0   0     0
 843.4   0.4    1.0e-1  0.2e-1  0       0       0       0       0       0       0    0   0    0   0     0
 883.1   0.2    3.0e-2  0.5e-2  0       0       0       0       0       0       0    0   0    0   0     0
 893.3   0.2    3.8e-2  0.8e-2  0       0       0       0       0       0       0    0   0    0   0     0
 894.8   0.2    8.8e-2  1.4e-2  0       0       0       0       0       0       0    0   0    0   0     0
****************************************************************************************************************
Upper Limits of Resonances
Note: enter partial width upper limit by chosing non-zero value for PT, where PT=<theta^2> for particles and...
Note: ...PT=<B> for g-rays [enter: "upper_limit 0.0"]; for each resonance: # upper limits < # open channels! 
Ecm     DEcm    Jr    G1        DG1   L1   PT      G2    DG2   L2  PT   G3   DG3   L3  PT  Exf  Int
   6.0  0.6     4.5   1.31e-61  0.0   0    0.0045  0.2   0.1   1   0    0    0     0   0   0.0  0
  68.3  0.7     2.5   6.02e-14  0.0   2    0.0045  0.2   0.1   1   0    0    0     0   0   0.0  0
  94.0  3.0     4.5   1.55e-8   0.0   0    0.0045  0.2   0.1   1   0    0    0     0   0   0.0  0
 126.7  0.8     4.5   9.41e-9   0.0   0    0.0045  0.2   0.1   1   0    0    0     0   0   0.0  0
 230.8  0.9     2.5   1.70e-6   0.0   2    0.0045  0.2   0.1   1   0    0    0     0   0   0.0  0
****************************************************************************************************************
Interference between Resonances [numerical integration only]
Note: + for positive, - for negative interference; +- if interference sign is unknown
Ecm     DEcm    Jr    G1       DG1   L1   PT     G2    DG2   L2  PT   G3   DG3   L3  PT  Exf  
!+- 
0.0     0.0     0.0   0.0      0.0   0    0      0.0   0.0   0   0    0.0  0.0   0   0   0.0  
0.0     0.0     0.0   0.0      0.0   0    0      0.0   0.0   0   0    0.0  0.0   0   0   0.0  
****************************************************************************************************************
Comments:
1. Gg=0.2 eV for upper limit resonances is a guess (not important since Gp<<Gg). 
2. Jp=4.5 for upper limit resonances assumed if no other information was available (values not uniquely known; 
   assume s-wave resonances).
3. Er<500 keV: with few exceptions, resonance energies are calculated from excitation energies of Lotay et al. 
   2009.
4. For proton partial widths of upper limit resonances C2S=1 assumed, with two exceptions: (i) C2S<0.002 for
   Er=127 keV (from Vogelaar et al. 1996); (ii) for Er=231 keV proton width upper limit is derived from 
   measured upper limit of resonance strength (Vogelaar 1989).
5. Er=68 keV: according to Lotay et al. 2009, the smallest possible orbital angular momentum is l=2.
\end{verbatim}
\normalsize
\vspace{5mm}
References: Lotay et al. \cite{Lot09}; Vogelaar \cite{Vog89}; Vogelaar et al. \cite{Vog96}.
\clearpage
\footnotesize

\normalsize
\vspace{5mm}
References: Chronidou et al. \cite{Chr99}; Endt \cite{End98}; Harissopulos et al. \cite{Har00}; Iliadis et al. \cite{Ili01}.
\clearpage
\footnotesize

\normalsize
\vspace{5mm}
References: Champagne et al. \cite{Cha88}; Endt \cite{End98}; Endt and Booten \cite{EnB93}; Iliadis et al. \cite{Ili01}; Timmermann et al. \cite{Tim88}.
\clearpage
\footnotesize
\begin{verbatim}
26Si(p,g)27P
****************************************************************************************************************
1               ! Zproj
14              ! Ztarget
0               ! Zexitparticle (=0 when only 2 channels open)
1.0078          ! Aproj         
25.9923         ! Atarget
0               ! Aexitparticle (=0 when only 2 channels open)
0.5             ! Jproj
0.0             ! Jtarget
0.0             ! Jexitparticle (=0 when only 2 channels open)
861.0           ! projectile separation energy (keV)
0.0             ! exit particle separation energy (=0 when only 2 channels open)
1.25            ! Radius parameter R0 (fm)
2               ! Gamma-ray channel number (=2 if ejectile is a g-ray; =3 otherwise)
****************************************************************************************************************
1.0             ! Minimum energy for numerical integration (keV)
7000            ! Number of random samples (>5000 for better statistics)
0               ! =0 for rate output at all temperatures; =NT for rate output at selected temperatures
****************************************************************************************************************
Nonresonant contribution
S(keVb) S'(b)           S''(b/keV)      fracErr Cutoff Energy (keV)
5.45e1  -1.66e-2        4.00e-6         0.4     4100.0
0.0     0.0             0.0             0.0     0.0
****************************************************************************************************************
Resonant Contribution
Note: G1 = entrance channel, G2 = exit channel, G3 = spectator channel !! Ecm, Exf in (keV); wg, Gx in (eV) !!
Note: if Er<0, theta^2=C2S*theta_sp^2 must be entered instead of entrance channel partial width
Ecm     DEcm    wg      Dwg     J       G1      DG1      L1    G2      DG2      L2   G3  DG3  L3 Exf    Int
 259    28      0       0       1.5     1.8e-4  0.72e-4  2     3.4e-3  1.7e-3   1    0   0    0  0.0    1
 772    33      0       0       2.5     4.3     1.7      2     3.3e-4  1.7e-4   2    0   0    0  0.0    1
1090    100     0       0       2.5     4.4     1.8      2     6.0e-4  3.0e-4   2    0   0    0  0.0    1
****************************************************************************************************************
Upper Limits of Resonances
Note: enter partial width upper limit by chosing non-zero value for PT, where PT=<theta^2> for particles and...
Note: ...PT=<B> for g-rays [enter: "upper_limit 0.0"]; for each resonance: # upper limits < # open channels!  
Ecm     DEcm    Jr    G1    DG1   L1   PT    G2    DG2    L2  PT   G3    DG3    L3  PT   Exf  Int
!0.0    0.0     0.0   0.0   0.0   0    0     0.0   0.0    0   0    0.0   0.0    0   0    0.0  0  
****************************************************************************************************************
Interference between Resonances [numerical integration only]
Note: + for positive, - for negative interference; +- if interference sign is unknown
Ecm     DEcm    Jr    G1    DG1   L1   PT    G2    DG2    L2  PT   G3    DG3    L3  PT   Exf  
!+- 
0.0     0.0     0.0   0.0   0.0   0    0     0.0   0.0    0   0    0.0   0.0    0   0    0.0  
0.0     0.0     0.0   0.0   0.0   0    0     0.0   0.0    0   0    0.0   0.0    0   0    0.0  
****************************************************************************************************************
Comments:
1. S-factor for direct capture to ground state calculated with C2S=0.58 from 27Mg mirror state.
2. The next resonances are expected at energies in excess of Er=2 MeV (see Moon et al. 2005).
\end{verbatim}
\normalsize
\vspace{5mm}
References: Moon et al. \cite{Moo05}.
\clearpage
\footnotesize

\normalsize
\vspace{5mm}
\clearpage
\footnotesize

\normalsize
\vspace{5mm}
\clearpage
\footnotesize

\normalsize
\vspace{5mm}
References: Endt \cite{End98}.
\clearpage
\footnotesize

\normalsize
\vspace{5mm}
References: Endt \cite{End90}.
\clearpage
\footnotesize
\begin{verbatim}
27P(p,g)28S
****************************************************************************************************************
1               ! Zproj
15              ! Ztarget
0               ! Zexitparticle (=0 when only 2 channels open)
1.0078          ! Aproj         
26.999          ! Atarget
0               ! Aexitparticle (=0 when only 2 channels open)
0.5             ! Jproj
0.5             ! Jtarget
0.0             ! Jexitparticle (=0 when only 2 channels open)
2460.0          ! projectile separation energy (keV)
0.0             ! exit particle separation energy (=0 when only 2 channels open)
1.25            ! Radius parameter R0 (fm)
2               ! Gamma-ray channel number (=2 if ejectile is a g-ray; =3 otherwise)
****************************************************************************************************************
1.0             ! Minimum energy for numerical integration (keV)
7000            ! Number of random samples (>5000 for better statistics)
0               ! =0 for rate output at all temperatures; =NT for rate output at selected temperatures
****************************************************************************************************************
Nonresonant contribution
S(keVb) S'(b)   S''(b/keV)      fracErr Cutoff Energy (keV)
2.00e1  0.0     0.0             0.5     10000.0
0.0     0.0     0.0             0.0     0.0
****************************************************************************************************************
Resonant Contribution
Note: G1 = entrance channel, G2 = exit channel, G3 = spectator channel !! Ecm, Exf in (keV); wg, Gx in (eV) !!
Note: if Er<0, theta^2=C2S*theta_sp^2 must be entered instead of entrance channel partial width
Ecm     DEcm    wg      Dwg     J       G1      DG1      L1    G2      DG2      L2   G3  DG3  L3 Exf    Int
1100    100     0       0       0       8.8e3   4.4e3    0     3.4e-4  1.7e-4   2    0   0    0  1518.0 1
****************************************************************************************************************
Upper Limits of Resonances
Note: enter partial width upper limit by chosing non-zero value for PT, where PT=<theta^2> for particles and...
Note: ...PT=<B> for g-rays [enter: "upper_limit 0.0"]; for each resonance: # upper limits < # open channels!  
Ecm     DEcm    Jr    G1    DG1   L1   PT    G2    DG2    L2  PT   G3    DG3    L3  PT   Exf  Int
!0.0    0.0     0.0   0.0   0.0   0    0     0.0   0.0    0   0    0.0   0.0    0   0    0.0  0  
****************************************************************************************************************
Interference between Resonances [numerical integration only]
Note: + for positive, - for negative interference; +- if interference sign is unknown
Ecm     DEcm    Jr    G1    DG1   L1   PT    G2    DG2    L2  PT   G3    DG3    L3  PT   Exf  
!+- 
0.0     0.0     0.0   0.0   0.0   0    0     0.0   0.0    0   0    0.0   0.0    0   0    0.0  
0.0     0.0     0.0   0.0   0.0   0    0     0.0   0.0    0   0    0.0   0.0    0   0    0.0  
****************************************************************************************************************
Comments:
1. Uncertainty of 100 keV for Er=1100 keV is a rough estimate only.
\end{verbatim}
\normalsize
\vspace{5mm}
\clearpage
\footnotesize
\begin{verbatim}
29P(p,g)30S
****************************************************************************************************************
1               ! Zproj
15              ! Ztarget
0               ! Zexitparticle (=0 when only 2 channels open)
1.0078          ! Aproj         
28.982          ! Atarget
0               ! Aexitparticle (=0 when only 2 channels open)
0.5             ! Jproj
0.5             ! Jtarget
0.0             ! Jexitparticle (=0 when only 2 channels open)
4399.0          ! projectile separation energy (keV)
0.0             ! exit particle separation energy (=0 when only 2 channels open)
1.25            ! Radius parameter R0 (fm)
2               ! Gamma-ray channel number (=2 if ejectile is a g-ray; =3 otherwise)
****************************************************************************************************************
1.0             ! Minimum energy for numerical integration (keV)
5000            ! Number of random samples (>5000 for better statistics)
0               ! =0 for rate output at all temperatures; =NT for rate output at selected temperatures
****************************************************************************************************************
Nonresonant contribution
S(keVb) S'(b)           S''(b/keV)      fracErr Cutoff Energy (keV)
7.3e1   -1.25e-2        0.0             0.4     1000.0
0.0     0.0             0.0             0.0     0.0
****************************************************************************************************************
Resonant Contribution
Note: G1 = entrance channel, G2 = exit channel, G3 = spectator channel !! Ecm, Exf in (keV); wg, Gx in (eV) !!
Note: if Er<0, theta^2=C2S*theta_sp^2 must be entered instead of entrance channel partial width
Ecm      DEcm    wg      Dwg    J     G1        DG1      L1    G2      DG2      L2   G3  DG3  L3 Exf    Int
 305.0   6.0     0       0      3     2.7e-5    1.1e-5   2     4.9e-3  2.5e-3   1    0   0    0  0.0    1
 489.0   40.0    0       0      2     2.2e-2    0.9e-2   2     4.2e-3  2.1e-3   1    0   0    0  0.0    1
 737.0   4.0     0       0      3     2.2e-1    0.9e-1   2     8.2e-3  4.1e-3   1    0   0    0  0.0    1
 818.4   3.1     0       0      3     3.6e-1    1.4e-1   3     9.4e-3  4.7e-3   1    0   0    0  0.0    1
 990.0   4.0     0       0      2     6.3e0     2.5e0    2     0.01    5.0e-3   1    0   0    0  0.0    0
****************************************************************************************************************
Upper Limits of Resonances
Note: enter partial width upper limit by chosing non-zero value for PT, where PT=<theta^2> for particles and...
Note: ...PT=<B> for g-rays [enter: "upper_limit 0.0"]; for each resonance: # upper limits < # open channels!  
Ecm     DEcm    Jr   G1      DG1   L1   PT      G2      DG2     L2  PT   G3    DG3    L3  PT   Exf  Int
 769.0   7.0    4    3.5e-4  0.0   4    0.0045  5.5e-3  2.8e-3  1   0    0     0      0   0    0.0  1
 769.0   7.0    0    1.1e1   0.0   0    0.0045  7.7e-3  3.9e-3  1   0    0     0      0   0    0.0  1
1443.0   5.0    4    1.0e-1  0.0   4    0.0045  2.7e-2  1.4e-2  1   0    0     0      0   0    0.0  1
****************************************************************************************************************
Interference between Resonances [numerical integration only]
Note: + for positive, - for negative interference; +- if interference sign is unknown
Ecm     DEcm    Jr    G1    DG1   L1   PT    G2    DG2    L2  PT   G3    DG3    L3  PT   Exf  
!+- 
0.0     0.0     0.0   0.0   0.0   0    0     0.0   0.0    0   0    0.0   0.0    0   0    0.0  
0.0     0.0     0.0   0.0   0.0   0    0     0.0   0.0    0   0    0.0   0.0    0   0    0.0  
****************************************************************************************************************
Comments:
1. Resonance energies calculated from excitation energies (Endt 1990 and Tab. I of Bardayan et al. 2007) and
   Q-value, except for Er=489 keV, which is based on the IMME (see Iliadis et al. 2001).
2. The spin-parity assignments are not unambiguous; they are based on experimental Jp restrictions (Endt 1990,
   Bardayan et al. 2007) and application of the IMME (Iliadis et al. 2001). 
3. Proton partial widths are computed using C2S of mirror states from (d,p) transfer experiment (Mackh et al. 
   1973).
4. Gamma-ray partial widths are computed from measured lifetimes of 30Si mirror states, except for Er=990 keV
   for which Gg=0.012 eV is a rough estimate (application of RULs to the known g-ray branchings of the mirror 
   state yields Gg<0.1 eV for this level).
\end{verbatim}
\normalsize
\vspace{5mm}
References: Bardayan et al. \cite{Bar07}; Endt \cite{End90}; Iliadis et al. \cite{Ili01}; Makh et al \cite{Mac73}.
\clearpage
\footnotesize

\normalsize
\vspace{5mm}
References: Endt \cite{End98}; Iliadis et al. \cite{Ili93}; Kalifa et al. \cite{Kal78}.
\clearpage
\footnotesize
\begin{verbatim}
31P(p,a)28Si
****************************************************************************************************************
1               ! Zproj
15              ! Ztarget
2               ! Zexitparticle (=0 when only 2 channels open)
1.0078          ! Aproj         
30.9737         ! Atarget
4.0026          ! Aexitparticle (=0 when only 2 channels open)
0.5             ! Jproj
0.5             ! Jtarget
0.0             ! Jexitparticle (=0 when only 2 channels open)
8863.78         ! projectile separation energy (keV)
6947.82         ! exit particle separation energy (=0 when only 2 channels open)
1.25            ! Radius parameter R0 (fm)
3               ! Gamma-ray channel number (=2 if ejectile is a g-ray; =3 otherwise)
****************************************************************************************************************
1.0             ! Minimum energy for numerical integration (keV)
5000            ! Number of random samples (>5000 for better statistics)
0               ! =0 for rate output at all temperatures; =NT for rate output at selected temperatures
****************************************************************************************************************
Nonresonant contribution
S(keVb) S'(b)           S''(b/keV)      fracErr Cutoff Energy (keV)
0.0     0.0             0.0             0.0     0.0
0.0     0.0             0.0             0.0     0.0
****************************************************************************************************************
Resonant Contribution
Note: G1 = entrance channel, G2 = exit channel, G3 = spectator channel !! Ecm, Exf in (keV); wg, Gx in (eV) !!
Note: if Er<0, theta^2=C2S*theta_sp^2 must be entered instead of entrance channel partial width
Ecm      DEcm   wg       Dwg      Jr    G1      DG1     L1   G2      DG2     L2   G3     DG3    L3  Exf   Int
 159.2   2.0    0        0        3     1.4e-11 0.6e-11 3    0.012   0.003   3    0.024  0.009  1   0.0   1
 599.4   1.0    2.5e-2   6.3e-3   0     0       0       0    0       0       0    0      0      0   0.0   0
 622.2   0.7    1.35     0.33     0     0       0       0    0       0       0    0      0      0   0.0   0
 846.8   0.5    1.40     0.35     0     0       0       0    0       0       0    0      0      0   0.0   0
 952.8   1.0    0.55     0.14     0     0       0       0    0       0       0    0      0      0   0.0   0
 984.0   3.0    14.0     1.5      0     0       0       0    0       0       0    0      0      0   0.0   0
1118.7   0.6    4.75     0.50     0     0       0       0    0       0       0    0      0      0   0.0   0
1358.7   0.8    20.0     2.0      0     0       0       0    0       0       0    0      0      0   0.0   0
1423.7   1.5    8.5      1.0      0     0       0       0    0       0       0    0      0      0   0.0   0
1428.8   1.5    13.3     1.5      0     0       0       0    0       0       0    0      0      0   0.0   0
1591.2   3.0    87.5     10.0     0     0       0       0    0       0       0    0      0      0   0.0   0
1662.9   3.0    23.2     2.5      0     0       0       0    0       0       0    0      0      0   0.0   0
1760.7   3.0    22.5     8.8      0     0       0       0    0       0       0    0      0      0   0.0   0
1836.3   1.0    2200.0   225.0    0     0       0       0    0       0       0    0      0      0   0.0   0
1914.8   1.0    100.0    10.0     0     0       0       0    0       0       0    0      0      0   0.0   0
1922.7   1.0    11.5     1.3      0     0       0       0    0       0       0    0      0      0   0.0   0
1928.2   1.0    52.5     5.0      0     0       0       0    0       0       0    0      0      0   0.0   0
1961.5   1.0    1650.0   175.0    0     0       0       0    0       0       0    0      0      0   0.0   0
1962.8   1.0    97.5     10.0     0     0       0       0    0       0       0    0      0      0   0.0   0
2051.3   3.0    275.0    28.0     0     0       0       0    0       0       0    0      0      0   0.0   0
2186.9   3.0    205.0    22.5     0     0       0       0    0       0       0    0      0      0   0.0   0
2360.2   3.0    21.8     2.3      0     0       0       0    0       0       0    0      0      0   0.0   0
2370.0   3.0    525.0    50.0     0     0       0       0    0       0       0    0      0      0   0.0   0
2582.0   3.0    135.0    15.0     0     0       0       0    0       0       0    0      0      0   0.0   0
2621.7   3.0    72.5     7.5      0     0       0       0    0       0       0    0      0      0   0.0   0
2719.5   3.0    160.0    17.5     0     0       0       0    0       0       0    0      0      0   0.0   0
2722.5   3.0    82.5     10.0     0     0       0       0    0       0       0    0      0      0   0.0   0
2739.9   3.0    13.0     1.3      0     0       0       0    0       0       0    0      0      0   0.0   0
2742.8   3.0    2.5      0.3      0     0       0       0    0       0       0    0      0      0   0.0   0
2744.7   3.0    925.0    175.0    0     0       0       0    0       0       0    0      0      0   0.0   0
2765.1   3.0    1250.0   250.0    0     0       0       0    0       0       0    0      0      0   0.0   0
2814.5   3.0    142.5    15.0     0     0       0       0    0       0       0    0      0      0   0.0   0
2857.1   3.0    62.5     7.5      0     0       0       0    0       0       0    0      0      0   0.0   0
2870.6   3.0    950.0    100.0    0     0       0       0    0       0       0    0      0      0   0.0   0
2940.4   3.0    35.0     7.5      0     0       0       0    0       0       0    0      0      0   0.0   0
2942.3   3.0    2500.0   500.0    0     0       0       0    0       0       0    0      0      0   0.0   0
****************************************************************************************************************
Upper Limits of Resonances
Note: enter partial width upper limit by chosing non-zero value for PT, where PT=<theta^2> for particles and...
Note: ...PT=<B> for g-rays [enter: "upper_limit 0.0"]; for each resonance: # upper limits < # open channels!  
Ecm     DEcm    Jr    G1      DG1     L1   PT       G2      DG2    L2  PT     G3     DG3    L3  PT   Exf  Int
195.2   2.0     1     6.4e-7  2.1e-7  1    0        1.2e-3  0.0    1   0.010  0.5    0.25   1   0    0.0  0
201.2   2.0     4     1.0e-9  0.0     4    0.0045   0.1     0.05   4   0      6.3e-3 1.3e-3 1   0    0.0  1
391.2   2.0     2     3.6e-4  0.6e-4  2    0        0.08    0.0    2   0.010  1.0    0.5    1   0    0.0  0
****************************************************************************************************************
Interference between Resonances [numerical integration only]
Note: + for positive, - for negative interference; +- if interference sign is unknown
Ecm     DEcm    Jr    G1      DG1     L1   PT   G2      DG2    L2  PT   G3    DG3    L3  PT   Exf  
+- 
 372.2  2.0     1     3.6e-3  1.0e-3  1    0    7.8     3.0    1   0    0.17  0.03   1   0    0.0   
1468.1  1.5     1     3800.0  600.0   1    0    2375.0  705.0  1   0    0.39  0.11   1   0    0.0     
****************************************************************************************************************
Comments:
1.  Resonance energies from Endt 1998.
2.  For measured resonance strengths assumed uncertainty of 25% if no value is listed in Endt 1998. 
3.  Assumed that doublet Ex=9023 (3-) + 9024 [6-(4-)] keV is dominated by former, natural parity, state.
4.  Gg=0.5 eV for Er=195 keV [Ex=9059 keV; Jp=(1,2)-] is a guess; furthermore, Jp=1- assumed since it better
    reproduces measured (p,g) resonance strength. 
5.  For Er=201 keV (4+): Gg<<Ga according to MacArthur et al. 1985; thus Gg=6.3e-3 eV is given by measured (a,g)
    resonance strength, while Ga is in the range of 0.03 - 0.2 eV.
6.  Level at Ex=9138 keV is disregarded since it originates from an unpublished study (see Endt 1998); it has 
    not been observed in proton transfer work.
7.  No indication that Ex=9170 keV level (Er=306 keV) decays by emission of alpha-particles (Ross et al. 1995); 
    disregarded the level here in agreement with unnatural parity assignment from the shell model (Endt 1998).
8.  For Er=391 keV, Gg=1 eV is a rough estimate; Ga/Gg<0.08 (Ross et al. 1995) yields then Ga<0.08 eV.
9.  Er=372 keV (Ex=9236 keV; 1-): partial widths are found from measured (p,g), (p,a) and (a,g) strengths; in
    particular, we disregard the other component in the doublet [Ex=9235 keV; (2+ to 5+)] for which the shell
    model assigns Jp=5+ (Endt 1998).
10. Er=1468 keV (Ex=10333 keV; Jp=1-): partial widths calculated from measured (p,g) and (p,a) strengths, 
   together with measured Gp (Kalifa et al. 1978). 
\end{verbatim}
\normalsize
\vspace{5mm}
References: Endt \cite{End98}; Kalifa et al. \cite{Kal78}; MacArthur et al. \cite{Mac85}; Ross et al. \cite{Ros95}.
\clearpage
\footnotesize
\begin{verbatim}
30S(p,g)31Cl
****************************************************************************************************************
1               ! Zproj
16              ! Ztarget
0               ! Zexitparticle (=0 when only 2 channels open)
1.0078          ! Aproj         
29.985          ! Atarget
0               ! Aexitparticle (=0 when only 2 channels open)
0.5             ! Jproj
0.0             ! Jtarget
0.0             ! Jexitparticle (=0 when only 2 channels open)
290.0           ! projectile separation energy (keV)
0.0             ! exit particle separation energy (=0 when only 2 channels open)
1.25            ! Radius parameter R0 (fm)
2               ! Gamma-ray channel number (=2 if ejectile is a g-ray; =3 otherwise)
****************************************************************************************************************
1.0             ! Minimum energy for numerical integration (keV)
7000            ! Number of random samples (>5000 for better statistics)
0               ! =0 for rate output at all temperatures; =NT for rate output at selected temperatures
****************************************************************************************************************
Non-resonant contribution
S(keVb) S'(b)   S''(b/keV)      fracErr Cutoff Energy (keV)
5.14e0  0.0     0.0             0.5     2000.0
0.0     0.0     0.0             0.0     0.0
****************************************************************************************************************
Resonant Contribution
Note: G1 = entrance channel, G2 = exit channel, G3 = spectator channel !! Ecm, Exf in (keV); wg, Gx in (eV) !!
Note: if Er<0, theta^2=C2S*theta_sp^2 must be entered instead of entrance channel partial width
Ecm     DEcm    wg      Dwg     J       G1      DG1      L1    G2      DG2      L2   G3  DG3  L3 Exf    Int
 461.0  15.0    0       0       0.5     7.7e-1  3.8e-1   0     8.6e-4  4.3e-4   1    0   0    0  0.0    1 
1463.0  2.0     0       0       2.5     2.70e1  1.35e1   2     1.0e-3  0.5e-3   1    0   0    0  0.0    1 
****************************************************************************************************************
Upper Limits of Resonances
Note: enter partial width upper limit by chosing non-zero value for PT, where PT=<theta^2> for particles and...
Note: ...PT=<B> for g-rays [enter: "upper_limit 0.0"]; for each resonance: # upper limits < # open channels!  
Ecm     DEcm    Jr    G1    DG1   L1   PT    G2    DG2    L2  PT   G3    DG3    L3  PT   Exf  Int
!0.0    0.0     0.0   0.0   0.0   0    0     0.0   0.0    0   0    0.0   0.0    0   0    0.0  0  
****************************************************************************************************************
Interference between Resonances [numerical integration only]
Note: + for positive, - for negative interference; +- if interference sign is unknown
Ecm     DEcm    Jr    G1    DG1   L1   PT    G2    DG2    L2  PT   G3    DG3    L3  PT   Exf  
!+- 
0.0     0.0     0.0   0.0   0.0   0    0     0.0   0.0    0   0    0.0   0.0    0   0    0.0  
0.0     0.0     0.0   0.0   0.0   0    0     0.0   0.0    0   0    0.0   0.0    0   0    0.0  
****************************************************************************************************************
Comments:
1. Er=461(15) keV energy from beta-delayed proton decay work of Axelsson et al. 1998.
2. Er=1463(2) keV energy from beta-delayed proton decay work of Fynbo et al. 2000.
3. See also Wrede et al. 2009.
\end{verbatim}
\normalsize
\vspace{5mm}
References: Axelsson et al. \cite{Axe98}; Fynbo et al. \cite{Fyn00}; Wrede et al. \cite{Wre09}.
\clearpage
\footnotesize

\normalsize
\vspace{5mm}
\clearpage
\footnotesize

\normalsize
\vspace{5mm}
\clearpage
\footnotesize
\begin{verbatim}
31Cl(p,g)32Ar
****************************************************************************************************************
1               ! Zproj
17              ! Ztarget
0               ! Zexitparticle (=0 when only 2 channels open)
1.0078          ! Aproj         
30.9924         ! Atarget
0               ! Aexitparticle (=0 when only 2 channels open)
0.5             ! Jproj
1.5             ! Jtarget
0.0             ! Jexitparticle (=0 when only 2 channels open)
2420.0          ! projectile separation energy (keV)
0.0             ! exit particle separation energy (=0 when only 2 channels open)
1.25            ! Radius parameter R0 (fm)
2               ! Gamma-ray channel number (=2 if ejectile is a g-ray; =3 otherwise)
****************************************************************************************************************
1.0             ! Minimum energy for numerical integration (keV)
7000            ! Number of random samples (>5000 for better statistics)
0               ! =0 for rate output at all temperatures; =NT for rate output at selected temperatures
****************************************************************************************************************
Nonresonant contribution
S(keVb) S'(b)   S''(b/keV)      fracErr Cutoff Energy (keV)
5.4e1   0.0     0.0             0.5     10000.0
0.0     0.0     0.0             0.0     0.0
****************************************************************************************************************
Resonant Contribution
Note: G1 = entrance channel, G2 = exit channel, G3 = spectator channel !! Ecm, Exf in (keV); wg, Gx in (eV) !!
Note: if Er<0, theta^2=C2S*theta_sp^2 must be entered instead of entrance channel partial width
Ecm     DEcm    wg      Dwg     J       G1      DG1      L1    G2      DG2      L2   G3  DG3  L3 Exf    Int
1600    100     0       0       2       4.2e3   2.1e3    0     1.1e-4  0.55e-4  2    0   0    0  0.0    1
****************************************************************************************************************
Upper Limits of Resonances
Note: enter partial width upper limit by chosing non-zero value for PT, where PT=<theta^2> for particles and...
Note: ...PT=<B> for g-rays [enter: "upper_limit 0.0"]; for each resonance: # upper limits < # open channels! 
Ecm     DEcm    Jr    G1    DG1   L1   PT    G2    DG2    L2  PT   G3    DG3    L3  PT   Exf  Int
!0.0    0.0     0.0   0.0   0.0   0    0     0.0   0.0    0   0    0.0   0.0    0   0    0.0  0  
****************************************************************************************************************
Interference between Resonances [numerical integration only]
Note: + for positive, - for negative interference; +- if interference sign is unknown
Ecm     DEcm    Jr    G1    DG1   L1   PT    G2    DG2    L2  PT   G3    DG3    L3  PT   Exf  
!+- 
0.0     0.0     0.0   0.0   0.0   0    0     0.0   0.0    0   0    0.0   0.0    0   0    0.0  
0.0     0.0     0.0   0.0   0.0   0    0     0.0   0.0    0   0    0.0   0.0    0   0    0.0  
****************************************************************************************************************
Comments:
1. Uncertainty of 100 keV for Er=1600 keV is a rough estimate only.
2. Dominant transition (E2) to ground state assumed for resonance, according to data on 32Si mirror (Endt and 
   van der Leun 1978).
\end{verbatim}
\normalsize
\vspace{5mm}
References: Endt and van der Leun \cite{End78}.
\clearpage
\footnotesize
\begin{verbatim}
32Cl(p,g)33Ar
****************************************************************************************************************
1               ! Zproj
17              ! Ztarget
0               ! Zexitparticle (=0 when only 2 channels open)
1.0078          ! Aproj         
31.9857         ! Atarget
0               ! Aexitparticle (=0 when only 2 channels open)
0.5             ! Jproj
1.0             ! Jtarget
0.0             ! Jexitparticle (=0 when only 2 channels open)
3343.0          ! projectile separation energy (keV)
0.0             ! exit particle separation energy (=0 when only 2 channels open)
1.25            ! Radius parameter R0 (fm)
2               ! Gamma-ray channel number (=2 if ejectile is a g-ray; =3 otherwise)
****************************************************************************************************************
1.0             ! Minimum energy for numerical integration (keV)
5000            ! Number of random samples (>5000 for better statistics)
0               ! =0 for rate output at all temperatures; =NT for rate output at selected temperatures
****************************************************************************************************************
Nonresonant contribution
S(keVb) S'(b)   S''(b/keV)      fracErr Cutoff Energy (keV)
3.25e1  0.0     0.0             0.5     2000.0
0.0     0.0     0.0             0.0     0.0
****************************************************************************************************************
Resonant Contribution
Note: G1 = entrance channel, G2 = exit channel, G3 = spectator channel !! Ecm, Exf in (keV); wg, Gx in (eV) !!
Note: if Er<0, theta^2=C2S*theta_sp^2 must be entered instead of entrance channel partial width
Ecm     DEcm    wg      Dwg     J       G1      DG1      L1    G2      DG2      L2   G3  DG3  L3 Exf    Int
  21    9       0       0       2.5     2.2e-43 1.1e-43  2     1.8e-2  0.9e-2   1    0   0    0  0      1
 113    9       0       0       3.5     5.2e-16 2.6e-16  2     1.9e-3  0.95e-3  1    0   0    0  0      1
 476    8       0       0       2.5     8.7e-4  4.35e-4  2     1.5e-2  0.75e-2  1    0   0    0  0      1
 847  100       0       0       0.5     3.8e1   1.9e1    0     1.5e-1  0.75e-1  1    0   0    0  0      1
1387  100       0       0       1.5     8.7e1   4.35e1   0     8.5e-2  4.25e-2  1    0   0    0  0      1
****************************************************************************************************************
Upper Limits of Resonances
Note: enter partial width upper limit by chosing non-zero value for PT, where PT=<theta^2> for particles and...
Note: ...PT=<B> for g-rays [enter: "upper_limit 0.0"]; for each resonance: # upper limits < # open channels!  
Ecm     DEcm    Jr    G1    DG1   L1   PT    G2    DG2    L2  PT   G3    DG3    L3  PT   Exf  Int
!0.0    0.0     0.0   0.0   0.0   0    0     0.0   0.0    0   0    0.0   0.0    0   0    0.0  0  
****************************************************************************************************************
Interference between Resonances [numerical integration only]
Note: + for positive, - for negative interference; +- if interference sign is unknown
Ecm     DEcm    Jr    G1    DG1   L1   PT    G2    DG2    L2  PT   G3    DG3    L3  PT   Exf  
!+- 
0.0     0.0     0.0   0.0   0.0   0    0     0.0   0.0    0   0    0.0   0.0    0   0    0.0  
0.0     0.0     0.0   0.0   0.0   0    0     0.0   0.0    0   0    0.0   0.0    0   0    0.0  
****************************************************************************************************************
Comments:
1. Direct capture S-factor calculated using shell-model spectroscopic factors (Schatz et al. 2005).
2. For Er=21, 113 and 476 keV the spin-parity assignments are not known experimentally (they are based on 
   Coulomb displacement energy calculations of Herndl et al. 1995).
3. The levels corresponding to Er=847 and 1387 keV are not known experimentally; the energies listed here are 
   based on Coulomb displacement energy calculations of Herndl et al. 1995; the uncertainty of 100 keV is a 
   rough guess only.
\end{verbatim}
\normalsize
\vspace{5mm}
References: Herndl et al. \cite{Her95}; Schatz et al. \cite{Sch05}.
\clearpage
\footnotesize

\normalsize
\vspace{5mm}
References: Johnson et al. \cite{Joh74}; Endt and van der Leun \cite{End78}; Iliadis et al. \cite{Ili94}; Iliadis et al. \cite{Ili01}; R\"opke et al. \cite{Roe02}; Ross et al \cite{Ros95}.
\clearpage
\footnotesize

\normalsize
\vspace{5mm}
References: Bosnjakovic et al. \cite{Bos68}; Endt and van der Leun \cite{End78}; Endt \cite{End98}; Iliadis et al. \cite{Ili94}; R\"opke et al. \cite{Roe02}; Ross et al \cite{Ros95}.
\clearpage
\footnotesize
\begin{verbatim}
34Ar(p,g)35K
****************************************************************************************************************
1               ! Zproj
18              ! Ztarget
0               ! Zexitparticle (=0 when only 2 channels open)
1.0078          ! Aproj         
33.9803         ! Atarget
0               ! Aexitparticle (=0 when only 2 channels open)
0.5             ! Jproj
0.0             ! Jtarget
0.0             ! Jexitparticle (=0 when only 2 channels open)
84.5            ! projectile separation energy (keV)
0.0             ! exit particle separation energy (=0 when only 2 channels open)
1.25            ! Radius parameter R0 (fm)
2               ! Gamma-ray channel number (=2 if ejectile is a g-ray; =3 otherwise)
****************************************************************************************************************
1.0             ! Minimum energy for numerical integration (keV)
7000            ! Number of random samples (>5000 for better statistics)
0               ! =0 for rate output at all temperatures; =NT for rate output at selected temperatures
****************************************************************************************************************
Nonresonant contribution
S(keVb) S'(b)   S''(b/keV)      fracErr Cutoff Energy (keV)
1.12e1  0.0     0.0             0.5     10000.0
0.0     0.0     0.0             0.0     0.0
****************************************************************************************************************
Resonant Contribution
Note: G1 = entrance channel, G2 = exit channel, G3 = spectator channel !! Ecm, Exf in (keV); wg, Gx in (eV) !!
Note: if Er<0, theta^2=C2S*theta_sp^2 must be entered instead of entrance channel partial width
Ecm     DEcm    wg      Dwg     J       G1      DG1      L1    G2      DG2      L2   G3  DG3  L3 Exf    Int
1469.0  5.0     0       0       0.5     2.2e3   1.1e3    0     2.1e-4  1.05e-4  1    0   0    0  0.0    1
****************************************************************************************************************
Upper Limits of Resonances
Note: enter partial width upper limit by chosing non-zero value for PT, where PT=<theta^2> for particles and...
Note: ...PT=<B> for g-rays [enter: "upper_limit 0.0"]; for each resonance: # upper limits < # open channels!  
Ecm     DEcm    Jr    G1    DG1   L1   PT    G2    DG2    L2  PT   G3    DG3    L3  PT   Exf  Int
!0.0    0.0     0.0   0.0   0.0   0    0     0.0   0.0    0   0    0.0   0.0    0   0    0.0  0  
****************************************************************************************************************
Interference between Resonances [numerical integration only]
Note: + for positive, - for negative interference; +- if interference sign is unknown
Ecm     DEcm    Jr    G1    DG1   L1   PT    G2    DG2    L2  PT   G3    DG3    L3  PT   Exf  
!+- 
0.0     0.0     0.0   0.0   0.0   0    0     0.0   0.0    0   0    0.0   0.0    0   0    0.0  
0.0     0.0     0.0   0.0   0.0   0    0     0.0   0.0    0   0    0.0   0.0    0   0    0.0  
****************************************************************************************************************
Comments:
1. Q-value from new 35K mass excess of Yazidjian et al. 2007.
2. Resonance energy from beta-delayed proton decay of 35Ca (Trinder et al. 1999). 
\end{verbatim}
\normalsize
\vspace{5mm}
References: Trinder et al. \cite{Tri99}; Yazidjian et al. \cite{Yaz07}. 
\clearpage
\footnotesize
\begin{verbatim}
35Ar(p,g)36K
****************************************************************************************************************
1               ! Zproj
18              ! Ztarget
0               ! Zexitparticle (=0 when only 2 channels open)
1.0078          ! Aproj         
34.975          ! Atarget
0               ! Aexitparticle (=0 when only 2 channels open)
0.5             ! Jproj
1.5             ! Jtarget
0.0             ! Jexitparticle (=0 when only 2 channels open)
1668.0          ! projectile separation energy (keV)
0.0             ! exit particle separation energy (=0 when only 2 channels open)
1.25            ! Radius parameter R0 (fm)
2               ! Gamma-ray channel number (=2 if ejectile is a g-ray; =3 otherwise)
****************************************************************************************************************
1.0             ! Minimum energy for numerical integration (keV)
5000            ! Number of random samples (>5000 for better statistics)
0               ! =0 for rate output at all temperatures; =NT for rate output at selected temperatures
****************************************************************************************************************
Nonresonant contribution
S(keVb) S'(b)           S''(b/keV)      fracErr Cutoff Energy (keV)
1.24e2  -8.66e-2        9.96e-5         0.4     1000.0
0.0     0.0             0.0             0.0     0.0
****************************************************************************************************************
Resonant Contribution
Note: G1 = entrance channel, G2 = exit channel, G3 = spectator channel !! Ecm, Exf in (keV); wg, Gx in (eV) !!
Note: if Er<0, theta^2=C2S*theta_sp^2 must be entered instead of entrance channel partial width
Ecm     DEcm    wg      Dwg     J     G1        DG1        L1    G2      DG2      L2   G3  DG3  L3  Exf   Int
-0.1    21      0       0       2     0.0372    0.0149     1     2.7e-4  1.35e-4  1    0   0    0   0.0   1
 224    21      0       0       2     5.7e-7    2.28e-7    0     1.0e-2  0.5e-2   1    0   0    0   0.0   1
 604    31      0       0       3     4.2e-1    1.68e-1    1     4.7e-4  2.35e-4  1    0   0    0   0.0   1
 744    31      0       0       2     2.5e0     1.0e0      0     1.1e-2  0.55e-2  1    0   0    0   0.0   1
****************************************************************************************************************
Upper Limits of Resonances
Note: enter partial width upper limit by chosing non-zero value for PT, where PT=<theta^2> for particles and...
Note: ...PT=<B> for g-rays [enter: "upper_limit 0.0"]; for each resonance: # upper limits < # open channels!  
Ecm     DEcm    Jr    G1    DG1   L1   PT    G2    DG2    L2  PT   G3    DG3    L3  PT   Exf  Int
!0.0    0.0     0.0   0.0   0.0   0    0     0.0   0.0    0   0    0.0   0.0    0   0    0.0  0  
****************************************************************************************************************
Interference between Resonances [numerical integration only]
Note: + for positive, - for negative interference; +- if interference sign is unknown
Ecm     DEcm    Jr    G1    DG1   L1   PT    G2    DG2    L2  PT   G3    DG3    L3  PT   Exf  
!+- 
0.0     0.0     0.0   0.0   0.0   0    0     0.0   0.0    0   0    0.0   0.0    0   0    0.0  
0.0     0.0     0.0   0.0   0.0   0    0     0.0   0.0    0   0    0.0   0.0    0   0    0.0  
****************************************************************************************************************
Comments:
1. Ficticious "Er=-0.1 keV" resonance is actually located at Er=2+-21 keV [using the latest Q-value]. Since this 
   low energy causes problems in the calculation of the Coulomb wave functions, the state is shifted just below 
   threshold.
\end{verbatim}
\normalsize
\vspace{5mm}
\clearpage
\footnotesize

\normalsize
\vspace{5mm} 
\clearpage
\footnotesize
\begin{verbatim}
35K(p,g)36Ca
****************************************************************************************************************
1               ! Zproj
19              ! Ztarget
0               ! Zexitparticle (=0 when only 2 channels open)
1.0078          ! Aproj         
34.988          ! Atarget
0               ! Aexitparticle (=0 when only 2 channels open)
0.5             ! Jproj
1.5             ! Jtarget
0.0             ! Jexitparticle (=0 when only 2 channels open)
2556.0          ! projectile separation energy (keV)
0.0             ! exit particle separation energy (=0 when only 2 channels open)
1.25            ! Radius parameter R0 (fm)
2               ! Gamma-ray channel number (=2 if ejectile is a g-ray; =3 otherwise)
****************************************************************************************************************
1.0             ! Minimum energy for numerical integration (keV)
7000            ! Number of random samples (>5000 for better statistics)
0               ! =0 for rate output at all temperatures; =NT for rate output at selected temperatures
****************************************************************************************************************
Nonresonant contribution
S(keVb) S'(b)   S''(b/keV)      fracErr Cutoff Energy (keV)
2.72e1  0.0     0.0             0.5     10000.0
0.0     0.0     0.0             0.0     0.0
****************************************************************************************************************
Resonant Contribution
Note: G1 = entrance channel, G2 = exit channel, G3 = spectator channel !! Ecm, Exf in (keV); wg, Gx in (eV) !!
Note: if Er<0, theta^2=C2S*theta_sp^2 must be entered instead of entrance channel partial width
Ecm     DEcm    wg      Dwg     J       G1      DG1      L1    G2      DG2      L2   G3  DG3  L3 Exf    Int
 459    43      0       0       2       1.1e-3  0.55e-3  0     3.8e-4  1.9e-4   2    0   0    0  0      1
****************************************************************************************************************
Upper Limits of Resonances
Note: enter partial width upper limit by chosing non-zero value for PT, where PT=<theta^2> for particles and...
Note: ...PT=<B> for g-rays [enter: "upper_limit 0.0"]; for each resonance: # upper limits < # open channels!  
Ecm     DEcm    Jr    G1    DG1   L1   PT    G2    DG2    L2  PT   G3    DG3    L3  PT   Exf  Int
!0.0    0.0     0.0   0.0   0.0   0    0     0.0   0.0    0   0    0.0   0.0    0   0    0.0  0  
****************************************************************************************************************
Interference between Resonances [numerical integration only]
Note: + for positive, - for negative interference; +- if interference sign is unknown
Ecm     DEcm    Jr    G1    DG1   L1   PT    G2    DG2    L2  PT   G3    DG3    L3  PT   Exf  
!+- 
0.0     0.0     0.0   0.0   0.0   0    0     0.0   0.0    0   0    0.0   0.0    0   0    0.0  
0.0     0.0     0.0   0.0   0.0   0    0     0.0   0.0    0   0    0.0   0.0    0   0    0.0  
****************************************************************************************************************
\end{verbatim}
\normalsize
\vspace{5mm}
\clearpage
\footnotesize

\normalsize
\vspace{5mm}
\clearpage
\footnotesize

\normalsize
\vspace{5mm}

\end{document}